\documentclass[fleqn,usenatbib,usedcolumn]{mnras}

\usepackage{graphicx}
\usepackage{multirow}
\usepackage{color}
\usepackage{amssymb}
\usepackage[fleqn]{amsmath}

\usepackage{siunitx}
\usepackage{comment} 
\usepackage{gensymb}
\usepackage{booktabs,tabularx}
\usepackage{flushend}
\usepackage{xspace}
\usepackage{physics}

\usepackage{nicefrac}
\usepackage{units}
\usepackage{verbatim}
\usepackage[shortlabels]{enumitem}
\usepackage{mathtools}
\usepackage{xcolor}
\usepackage{soul}

\usepackage{hyperref}

\title[Exponential shapelets]{Exponential shapelets:\\basis functions for data analysis of isolated features}

\author[Joel Berg\'e et al.]
{\parbox{\textwidth}{Joel Berg\'e,$^1$\thanks{e-mail: {\tt joel.berge@onera.fr}}
Richard Massey,$^{2}$ 
Quentin Baghi,$^3$ 
and Pierre Touboul$^4$}
\vspace{0.3cm}
\\$^1$DPHY, ONERA, Universit\'e Paris Saclay, F-92322 Ch\^atillon, France
\\$^2$Centre for Extragalactic Astronomy, Department of Physics, Durham University, Durham DH1 3LE, UK
\\$^3$NASA Goddard Space Flight Center, 8800 Greenbelt Rd, Greenbelt, MD 20771, USA
\\$^4$DSG, ONERA, Universit\'e Paris Saclay, Chemin de la Huni\`ere, F-91123 Palaiseau Cedex, France
}

\begin{document}

\maketitle

\label{firstpage}

\begin{abstract}
We introduce one- and two-dimensional `exponential shapelets': orthonormal basis functions that efficiently model isolated features in data. 
They are built from eigenfunctions of the quantum mechanical hydrogen atom, and inherit mathematics with elegant properties under Fourier transform, and hence (de)convolution. 
For a wide variety of data, exponential shapelets compress information better than Gauss-Hermite/Gauss-Laguerre (`shapelet') decomposition, and generalise previous attempts that were limited to 1D or circularly symmetric basis functions.
We discuss example applications in astronomy, fundamental physics and space geodesy. 
\end{abstract}

\begin{keywords}
Methods: Data analysis -- Physical data and processes
\end{keywords}

%%%%%%%%%%%%%%%%%%%%%%%%%%%%%%%%%%%%%%%%%%%%%%%%%%%%%%%%%%%%%%%%%%%%%%%%%%%%%
%% Main text
\section{Introduction}

A frequent task in data analysis is to categorise and quantify the shapes of localised objects -- such as transient events in a (one-dimensional) time-series, or regions of interest in a (two-dimensional) image. 
It is such a universal challenge that methods developed for one field frequently turn out to be useful in others.
For example, astrophysicists measure the shapes of distant galaxies by decomposing them into  orthogonal basis functions, such as CHEFs \citep{jimenezteja12} or (Gaussian) shapelets (\citealt{bernstein02,shapelets1, shapelets2, polar_shapelets}). 
Shapelets have been used to analyse data in other branches of astrophysics, modelling extrasolar planets \citep{exoplanets1,exoplanets2}, the distribution of dark matter \citep{birrer15,tagore16}, or flashes of pulsars \citep{pulsar1,pulsar2,pulsar3}.
They have also been used in 
medical imaging \citep{skin,breastcancer2}, 
pattern recognition in human vision \citep{v1fields}
or artificial vision \citep{pedestrians},
data compression \citep{datacompress},
and the manufacture of nanoscale thin films \citep{surfaceselfassemblyimaging,nanoscale}.

Gaussian shapelets are based on eigenfunctions of the quantum mechanics harmonic oscillator.
In one-dimensional form, they are Gauss-Hermite functions, which seem to be well adapted in several time-series analyses, especially when transients (whose shape is close to damped oscillations) must be detected, characterised and/or corrected for. This is the case for instance in fundamental physics for MICROSCOPE `crackles' (e.g. \citealt{baghi15,berge15}), space geodesy for GRACE `twangs' \citep{flury08, peterseim10, peterseim14b} or `glitches' in gravitational waves searches with LIGO (\citealt{cornish15, powell15, powell17, principe17} and references therein).
In two-dimensional form, they can be expressed equivalently as either Gauss-Hermite (Cartesian; \citealt{shapelets1}) or Gauss-Laguerre (polar; \citealt{bernstein02,polar_shapelets}) functions. Owing to their quantum mechanical origin, they have elegant properties (they make a complete set) under Fourier transform, and are hence efficient for operations involving convolution or deconvolution of two images.

The main limitation of shapelets is that they are perturbations around a Gaussian, which is flat near its peak. They inefficiently parameterise peaky features, including the distant galaxies for which they were originally suggested \citep{melchior10}. Galaxies have a two-dimensional surface density that decreases approximately exponentially with distance from the centre. 
Capturing the steep gradient near the centre requires a weighted sum of many Gaussian shapelet basis functions, which then overfit noise in the extended wings. Attempting to overcome this limitation, \cite{ngan09} developed `Sersiclets', although they were forced to be circularly symmetric, and have not seen wider applications. 

In this paper, we extend the quantum mechanical framework underpinning Gaussian shapelets, and define 1D and 2D {\it exponential shapelets} based on wavefunctions of the hydrogen atom.
These functions are perturbations around a decreasing exponential, and should efficiently model any arbitrarily-shaped but centrally-peaked regions of interest.
The perturbations themselves are Laguerre polynomials.

We introduce 1D exponential shapelets in Section~\ref{sect_1D_exponential_shapelets}, and 2D exponential shapelets in Section~\ref{sect_2D_exponential_shapelets}.
In both cases, we describe their main properties, compare them to Gaussian shapelets, and provide example uses.
We conclude in Section~\ref{sect_conclusion}.

\section{1D exponential shapelets} \label{sect_1D_exponential_shapelets}

\subsection{Definition}

The one-dimensional hydrogen atom is the solution to the motion of a particle in a 1D Coulomb potential $1/|x|$. 
In this paper, we will neither dwell on its rich history, nor the debates about the finitude of its ground state, and about the existence of even wavefunctions (see \citealt{nieto79, palma06, nunez11, nunez14} and references therein). 
Instead, we simply exploit the normalized 1D hydrogen atom wavefunctions as given by \cite{palma06} to define the 1D exponential shapelet basis functions 
\begin{equation} \label{eq_1D exponential shapelets}
\Psi_n^+(x; \beta) \! = \! \frac{(-1)^{n-1}}{\sqrt{n^3\beta}}\, \frac{2x}{n\beta}\, L_{n-1}^1\left( \frac{2x}{n\beta}\right) \exp\left(-\frac{x}{n\beta}\right)~~ \forall x \geqslant 0, 
\end{equation}
for $n\geqslant 1$, where $L_{n-1}^1$ are the generalized Laguerre polynomials (see e.g. \citealt{polar_shapelets}).
The characteristic scale size $\beta$ corresponds to the Bohr radius, and $n$ is 
the eigenfunction energy level.

Note that, contrary to normal procedure in quantum physics, we restrict 1D exponential shapelets on positive $x$, and refrain from defining the negative-$x$ part $\Psi_n^-(x; \beta) = -\Psi_n^+(-x; \beta)$ (for $x < 0$). Hence, we do not follow \cite{palma06} and do not define the even and odd wavefunctions $\Psi_n^+(x; \beta) \pm \Psi_n^-(x; \beta)$. 
Events in time series can often be adequately described by their behaviour after a certain moment (in this case, $x=0$). 
Henceforth, we shall therefore drop the $^+$ superscript, to write more simply $\Psi_n(x; \beta) \equiv \Psi_n^+(x; \beta)$.
These functions are continuously differentiable, smoothly departing from zero at $x=0$ and tending back to zero as $x\rightarrow\infty$.

Because the basis functions (\ref{eq_1D exponential shapelets}) are eigenfunctions of the one-dimensional hydrogen atom's Hamiltonian, they form an orthogonal basis of the square integrable functions $L^2([0,\infty[, \langle \cdot, \cdot \rangle)$ Hilbert space equiped with the inner product $\langle f(x),g(x) \rangle=\int_0^\infty f(x)g^*(x) {\rm d}x$ and an asterisk denotes complex conjugation.
They are orthonormal, in the sense that $\int_0^\infty \Psi_n(x; \beta)\Psi_m(x; \beta)\, {\rm d}x = \delta_{nm}$ where $\delta_{nm}$ is the Kronecker symbol, and complete, in the sense that $\sum_{n=1}^{\infty} \Psi_n(x; \beta) \Psi_n(x'; \beta) = \delta(x - x')$.

They thus form a basis, on which we can uniquely decompose a function as
\begin{equation} \label{eq_h1decomp}
f(x) = \sum_{n=1}^\infty f_n \Psi_n(x; \beta),
\end{equation}
with coefficients $f_n$ given by an overlap integral
\begin{equation} \label{eq_h1coeffs}
f_n = \int_0^\infty  f(x) \Psi_n(x; \beta)\, {\rm d}x.
\end{equation}
Bessel's inequality then assures us that for any function $f \in  L^2([0,\infty[, \langle \cdot, \cdot \rangle)$,
\begin{equation}
\sum_{n=1}^{\infty} |f_n|^2 \leqslant  ||f||^2,
\end{equation}
where $||.||$ is the L2 norm, so that the series (\ref{eq_h1decomp}) converges and coefficients $f_n$ must vanish as $n$ increases.
The series can therefore be truncated for suitably localised functions, at some value $n\leqslant n_{\rm max}$.

\begin{figure}
\centering
\includegraphics[width=0.45\textwidth,angle=0]{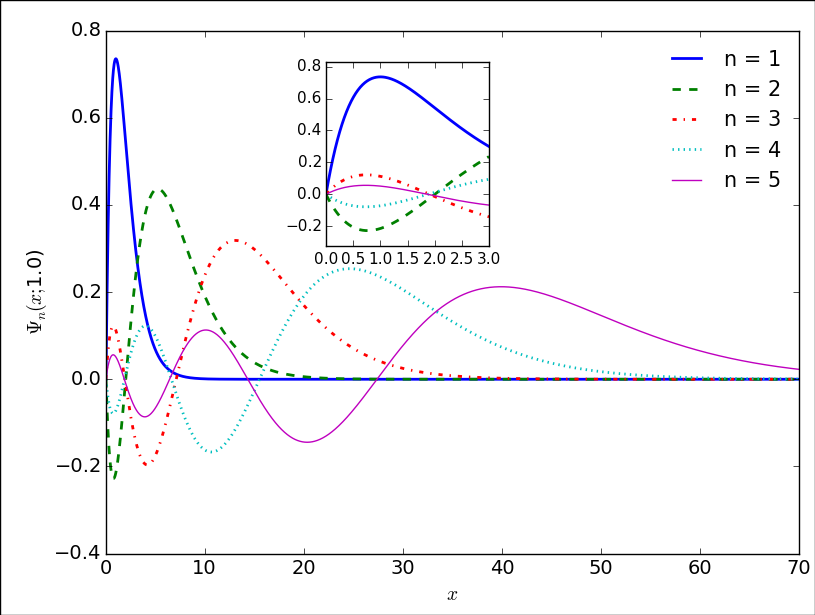}
\caption{\small The first few 1D exponential shapelets $\Psi_n(x)$, for $\beta\!=\!1$. The inset shows a zoom on the smallest $x$.} \label{fig_1D exponential shapelets}
\end{figure}

\subsection{Properties}

\subsubsection{Maximum and minimum effective scales}\label{sect_1D_theta_minmax}

The first few 1D exponential shapelet basis functions $\Psi_n(x; \beta)$ are shown in Fig.\ \ref{fig_1D exponential shapelets}.
As the order $n$ increases (with constant $\beta$), the largest-scale oscillation dominates, and rapidly acquires a larger extent.
However, the smallest oscillations always remain roughly the same size.

To act as a convenient figure-of-merit, we follow \cite{shapelets1} in defining the largest scale that can be described by an exponential shapelet model as
\begin{equation} 
\theta_{\rm max} \approx \sqrt{\frac{\int  x^2 \Psi_{n_{\rm max}}(x;\beta)\,{\rm d}x}{\int \Psi_{n_{\rm max}}(x;\beta)\,{\rm d}x}} = \beta\sqrt{2{n_{\rm max}}(2{n_{\rm max}}+1)}\,.
\end{equation}
Empirically, the smallest scale that can be described is
\begin{equation}
\theta_{\rm min} \approx \beta.
\end{equation}

As ${n_{\rm max}}\rightarrow\infty$, the range of scales modelled by a 1D exponential shapelet decomposition, $\theta_{\rm max}/\theta_{\rm min}\propto n_{\rm max}=n_{\rm coeffs}$.
However, the resolution of an exponential shapelet model is greatest near the origin, and decreases away from it.
This behaviour is very different from Gaussian shapelets, where the resolution is more spatially uniform: increasing $n$ increases the scale of the model slowly, while simultaneously adding smaller-scale oscillations (see Fig.\ 1 of \citealt{shapelets1}). 
The wide effective range of scales for exponential shapelets is what will allow them to efficiently describe spiky but long-duration events.

\subsubsection{Fourier and Laplace transforms}

To compute the Fourier transform of the 1D exponential shapelet basis functions, we adopt a convention where the Fourier transform is defined as
\begin{equation}
\begin{array}{ccc}
\tilde{f}(k) & = & (2\pi)^{-\frac{1}{2}} \int_{-\infty}^\infty  f(x) {\rm e}^{ikx}\, {\rm d}x, \\
f(x) & = & (2\pi)^{-\frac{1}{2}} \int_{-\infty}^\infty \tilde{f}(k) {\rm e}^{-ikx}\,{\rm d}k 
\end{array}
\end{equation}
for any function $f(x)$, where $i^2 = -1$. The Fourier transform of the basis function $\Psi_n(x; \beta)$ is then (setting $f(x)=0$ for $x<0$)
\begin{equation} \label{eq_h1_fourier}
\tilde{\Psi}_n(k; \beta) = (-1)^n \sqrt{\frac{2n\beta}{\pi}} \frac{(n\beta k -i)^{2n}}{[(n\beta k)^2 + 1]^{n+1}}\,.
\end{equation}
Note that the real part of the Fourier transform is even for all $n$, while its imaginary part is odd for all $n$. Its modulus is a Lorentzian function centered on $k=0$ whose width parameter $\Gamma\equiv2/(n\beta)$ is inversely proportional to $\beta$ (as typical for exponentially suppressed oscillations)
\begin{equation}
\left|\tilde{\Psi}_n(k; \beta)\right| = \sqrt{\frac{2\pi}{n\beta}} \frac{1}{\pi} \frac{\Gamma/2}{k^2 + (\Gamma/2)^2}\, .
\end{equation}

\begin{figure}
\centering
\includegraphics[width=0.5\textwidth,angle=0]{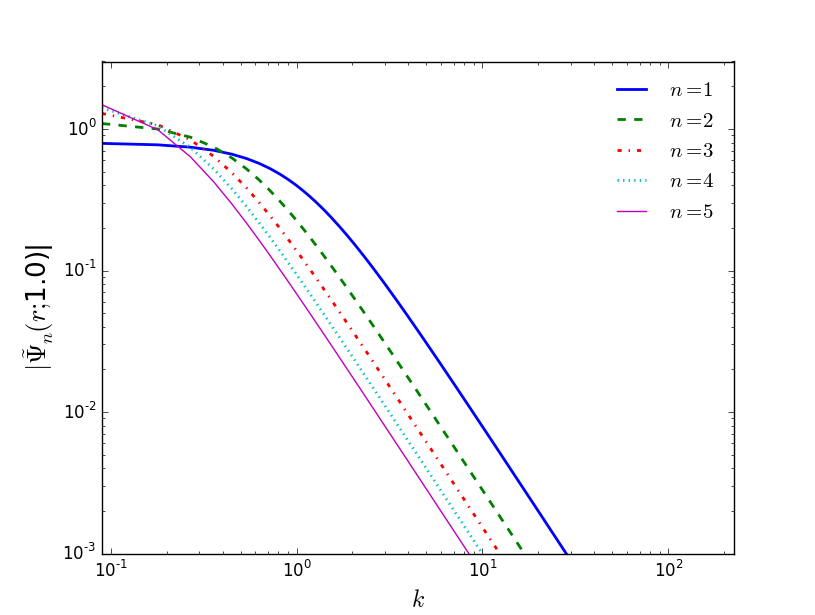}
\caption{\small Exponential shapelets are well localized in frequency space. This shows the modulus of the Fourier transform of the first few 1D exponential shapelets $\Psi_n(x;\beta)$, for $\beta=1$. 
\label{fig_1D_exponential_ shapelets_fourier}}
\end{figure}

With conventions similar to those for the Fourier transform, the Laplace transform is given by
\begin{equation} \label{eq_h1_laplace}
\mathcal{L}(\Psi_n(x; \beta))(s) = (-1)^{n-1} 2\sqrt{n\beta} \frac{(n\beta s-1)^{n-1}}{(n\beta s+1)^{n+1}}\, .
\end{equation}

Although Eqs.\ (\ref{eq_h1_fourier}) and (\ref{eq_h1_laplace}) are not as simple as for Gaussian shapelets (the Fourier transform of a Gaussian shapelet is itself a Gaussian shapelet), closed forms exist for exponential shapelets that are easy to implement.
The Fourier and Laplace transforms of 1D exponential shapelets are also well localized in frequency space (Fig.\ \ref{fig_1D_exponential_ shapelets_fourier}).

\subsubsection{Convolution}

Convolution is an inevitable operation during signal acquisition, whereby an input (`real') signal undergoes the measurement apparatus' transfer function. What is ultimately measured is the convolution of the input signal with the transfer function.

Let us consider the convolution of two functions $f(x)$ and $g(x)$, whose convolution product is
\begin{equation}
h(x) \equiv (f \star g)(x) \equiv \int_{-\infty}^\infty {\rm d}x' f(x-x')g(x').
\end{equation}
Following the same arguments as \citet{shapelets1}, 
the functions can all be decomposed into exponential shapelets, perhaps with different scale sizes $\alpha$, $\beta$ and $\gamma$.
We can then relate the 1D exponential shapelet space coefficients $h_{n;\gamma}$ to $f_{m;\beta}$ and $g_{l;\alpha}$ via
\begin{equation}
h_n = \sum_{m,l=1}^\infty C_{nml}f_m g_l,
\end{equation}
where the convolution tensor is given by
\begin{multline} \label{eq_cnml}
C_{nml}(\gamma, \alpha, \beta) = \sqrt{\frac{8\, \alpha\beta\gamma\, nml}{\pi} } 
 \sum_{u=0}^{2m}\sum_{v=0}^{2l}\sum_{w=0}^{2n} i^{u+v+w} \times \\ 
 \binom{2m}{u} \binom{2l}{v}\binom{2n}{w} (m\alpha)^u (l\beta)^v (n\gamma)^n \mathcal{I}_{u,v,w}^{m,l,n}
\end{multline}
and $\binom{n}{m}$ is the binomial coefficient. A proof of Eq.\ (\ref{eq_cnml}) is provided in Appendix \ref{sect_app1}.

The integral
\begin{equation} \label{eq_integral}
\mathcal{I}_{u,v,w}^{m,l,n} \!\! \equiv \!\!\!
\int_{-\infty}^\infty \!\!\frac{ (-1)^w ~ k^{u+v+w}~~{\rm d}k}{\left[(m\alpha k)^2 + 1\right]^{m+1} \left[(l\beta k)^2 + 1\right]^{l+1} \left[(n\gamma k)^2 + 1\right]^{n+1}}
\end{equation}
appearing in Eq.\ (\ref{eq_cnml}) is zero if $u+v+w$ is odd (a proof is provided in Appendix \ref{sect_app2}). Hence, $C_{nml}(\gamma,\alpha,\beta)$ is always wholly real. 
Under some conditions, it can be estimated analytically and expressed as a sum of converging series (see Eqs.\ \ref{eq_app_I}, \ref{eq_app_J} and \ref{eq_app_K} in Appendix \ref{sect_app3}). 
Those conditions are often not met, but the integrand is a well-behaved function, so the integration can also be performed numerically.
We find that numerical integration is most convenient if the function is cut into three segments (see Eqs.\ \ref{eq_app_Iint}, \ref{eq_app_Jint} and  \ref{eq_app_Kint} in Appendix \ref{sect_app3}).

\subsubsection{Rescaling}

1D exponential shapelets obey the integral property
\begin{equation} \label{eq_int_prop}
\int_0^\infty \Psi_n(x;\beta) {\rm d}x = 2\sqrt{n\beta}.
\end{equation}
Using this, it can be shown that under a rescaling $x \rightarrow x'=ax$, the coefficients $f_n$ of a model  (\ref{eq_h1coeffs}) transform as
\begin{equation}
f'_{n;\beta} = a^{-1/2} f_{n;a\beta}\, ,
 \end{equation} 
 and under $f(x) \rightarrow f'(x)=kf(x)$, the coefficients are themselves multiplied by $k$.

\subsection{Shape characterization}

Let $f(x) = \sum_n f_n \Psi_n(x; \beta)$ be a 1D object decomposed into 1D exponential shapelets. 
In terms of its shapelet coefficients, its integral (total `flux') is
\begin{equation}
F\equiv \int_{-\infty}^\infty f(x){\rm d}x = 2\sqrt{\beta} \sum_{n=1}^\infty \sqrt{n}\, f_n\, ,
\end{equation}
which can be readily shown from the integral property (\ref{eq_int_prop}).

Provided $F \neq 0$, its barycenter (centroid) is 
\begin{equation}
x_c  \equiv \frac{1}{F}\int_{-\infty}^\infty x f(x){\rm d}x =  \frac{4\sqrt{2}\beta^{3/2}}{F} \sum_{n=1}^\infty  n^2 f_n\,,
\end{equation}
and it has a characteristic  size 
\begin{equation} \label{eq_rms1}
r_{\rm c}^2 \equiv \frac{1}{F} \int_{-\infty}^\infty x^2 f(x){\rm d}x = \frac{4 \beta^{5/2}}{F} \sum_{n=1}^\infty n^3 (2n^2+1)\, f_n\, .
\end{equation}
This size can be used to determine the exponential decay rate of the object, for instance, when modelling damped oscillations. 

Note that series (\ref{eq_rms1}) converges only if the amplitudes of the 1D exponential shapelet coefficients decrease faster than $n^{-6}$, which may not always be the case. Care must therefore be taken to check for convergence when characterising the shape of a feature using this technique.

\subsection{Exponential shapelets modelling in practice} \label{ssect_focus}

As shown above (Eq. \ref{eq_h1decomp}), a 1D feature is straightforward to model for a given couple ($n_{\rm max}$, $\beta$), where $n_{\rm max}$ is the maximum order of the truncated sum (\ref{eq_h1decomp}). For example, a linear least-square method can be efficiently used for this purpose. Then the model depends non-linearly on the two parameters $n_{\rm max}$ and $\beta$ that can be optimised by iteratively minimizing the residuals between the observed feature and its model. This procedure was described at length, in the 2D case, by \cite{polar_shapelets}.

\subsection{Example applications}

This section demonstrates three possible applications of 1D exponential shapelets.
We start by modelling exponentially suppressed oscillations, which are measurements throughout experimental physics, including the response of accelerometers\footnote{http://www.onera.fr/en/dphy} onboard the space missions MICROSCOPE \citep{touboul17}, 
GRACE \citep{tapley04} and GOCE \citep{rummel11}.
We then discuss a potential application to unmodelled bursts in the analysis of gravitational waves.
Exponential shapelets may also be convenient to model charge transfer inefficiency trailing due to radiation damage in spacebourne imaging detectors \citep{cti}, although we do not explore that further here.

\subsubsection{Cleaning accelerometer time series data, ~~~~~~~~~~~~~~~~~~~ and modelling an experiment's response function}

\begin{figure}
\centering
\includegraphics[width=0.45\textwidth,angle=0]{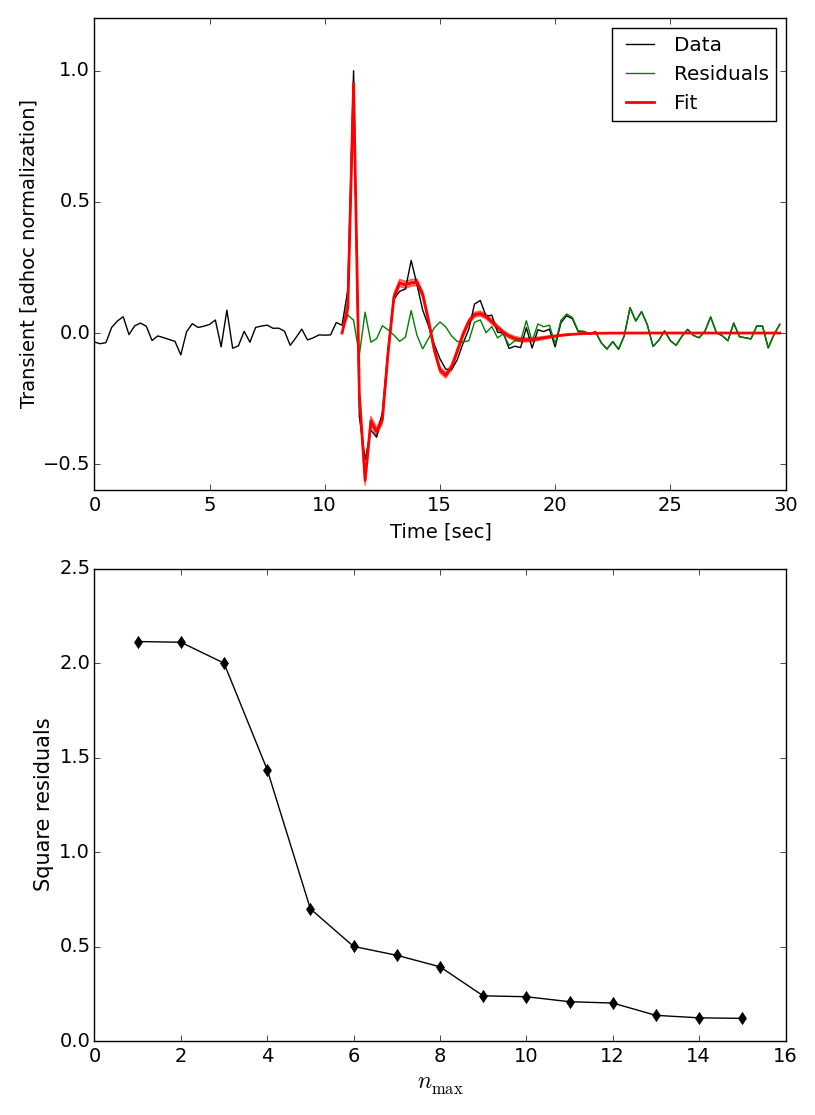}
\caption{\small Test data for 1D exponential shapelets. Upper panel: transient in MICROSCOPE accelerometer time series, as observed (black) and reconstructed from a 1D exponential model (red), with model residuals (green). Lower panel: convergence of the model as we increase the order of the model $n_{\rm max}$.} \label{fig_crtf}
\end{figure}

MICROSCOPE tested the Weak Equivalence Principle (WEP) by precisely measuring the differential acceleration experienced by two concentric cylindrical test-masses onboard a drag-free satellite in Low Earth Orbit\footnote{The mission ended on October 16, 2018; data analysis is still underway.}.
In theory, any non-zero difference at a well defined frequency $f_{\rm EP}$ (which depends on the orbit and attitude characteristics of the satellite) is a signature of the violation of the WEP. 

In practice, many transients are apparent in MICROSCOPE data (the upper panel of Fig.\ \ref{fig_crtf} shows a typical high-signal-to-noise example). These transients are generally caused by crackles of the satellite's coating (because of temperature variations), crackles of the satellite's gas tanks (as their pressure decreases as the gas is consumed), or impacts with micro-meteorites. Such transients occupy frequencies higher than $f_{\rm EP}$, so they do not directly impact a possible WEP violation signal. However, it is necessary to detect and mask them in measured time-series, then either reconstruct the corresponding `missing data' \citep{berge15, pires16} to allow for a least square fit of the expected WEP violation signal in the frequency domain \citep{touboul17}, or adapt the maximum likelihood technique to take missing data into account \citep{baghi15, baghi16}. Existing techniques are suboptimal, as they may affect the noise characteristics. Moreover, transients could be considered as conveying useful information. They are created by an external impulse; if this is assumed to be instantaneous (Dirac), the  shape and relaxation time of the observed signal is a measurement of MICROSCOPE's transfer function.

1D exponential shapelets are well-matched to these transient signals. The upper panel of Fig.\ \ref{fig_crtf} shows a transient in the time domain: the observed data are shown in black, a model (fitted between $t=10$s and $t=25$s) and its 68\% confidence interval are shown in red, and residuals consistent with noise are shown in green. This model uses $n_{\rm max}=15$, meaning that it has compressed 60 data points in the time domain description (15 seconds sampled at 4Hz) into 15 shapelet coefficients. Nonetheless, the model has rich, empirical freedom to capture multiple response modes of the instrument's complex structure. The lower panel of Fig. \ref{fig_crtf} shows the convergence of the model as we increase $n_{\rm max}$, quantified as the square residuals between the observed transient and the model. Most of the information is contained in coefficients $3\leqslant n \leqslant13$.

An extension to this process will be presented in a future paper. By fitting 1D exponential shapelet coefficients to many transients, it is possible to model temporal variations in MICROSCOPE's instrument's relaxation time via parameter $\beta$ or the exponential envelope scale $r_{\rm rms}$. Interpolating these models then yields a model of the transfer function at any time, to either gain insight into the instrument's performance, or to correct (deconvolve) signals with an `inverse transfer function' transform. 
Note that this procedure is similar to techniques developed for 2D astronomical image processing, where a Point Spread Function (PSF) is determined from the shape of stars then interpolated across the data (see e.g.\ \citealt{berge12, gentile13, kilbinger15}).

\subsubsection{Characterisation of perturbing signals in space-borne geodesy missions}

The GRACE mission \citep{tapley04} revolutionised geodesy by measuring the Earth's gravitational field with unprecedented precision. 
Two satellites followed each other on the same orbit, monitoring the distance between them via microwave ranging. 
In theory, any variation in their relative speed or distance can be ascribed to variations in the Earth's geopotential.

In practice, the satellites were also subject to non-gravitational forces.
These were measured by ultrasensitive accelerometers, for removal during post-processing \citep{flury08,peterseim10}.
\cite{peterseim14a} and \cite{peterseim14b} modelled transients (known as `spikes') in accelerometer data using a piecewise function made from the derivative of a Gaussian, a 3rd order polynomial and a damped oscillation.
Some were successfully classified as due to either the satellite's heaters, or activation of its magnetic torquers -- but no physical origin could be assigned to others, known as `twangs'.

Tentative correlations of twangs with the position of the satellite along its orbit hint at a possible geophysical origin.
Both categories of twangs are compactly-represented as 1D exponential shapelets, so we will investigate in a future paper whether these provide a cleaner set of shape parameters to categorise and understand their origin.

\subsubsection{Unmodelled bursts and glitches in gravitational waves data analysis}

A wealth of methods have been developed to search for, characterise and classify unmodelled bursts and instrumental glitches in searches for gravitational waves \citep[see e.g.][and references therein]{cornish15, powell15, powell17, principe17}.
Glitches are often modelled using `Sine-Gaussian waveforms' \citep{principe17}.
These have similar properties to Gaussian shapelets, although shapelets can encode more details.
1D exponential shapelets could further optimise the data compression of complex glitch shapes that often exhibit damped oscillations.

Shapelets might therefore improve the detection, characterisation and classification of instrumental glitches in gravitational wave detectors. Indeed, bursts and glitches are usually detected in the time-frequency domain, which is easily reproduced in Gaussian shapelets-time space. If 1D exponential shapelets more efficiently model the information in a glitch, exponential shapelet-time space would be even better. 
We will investigate in a future paper whether glitches can be detected by scanning a matched-filter, and correlating the measured signal with a 1D exponential shapelet model, while leaving $\beta$ (and possibly $n_{\rm max}$) free.

\subsection{Comparison with Gaussian shapelets}

\begin{figure}[t]
\centering
\includegraphics[width=0.45\textwidth,angle=0]{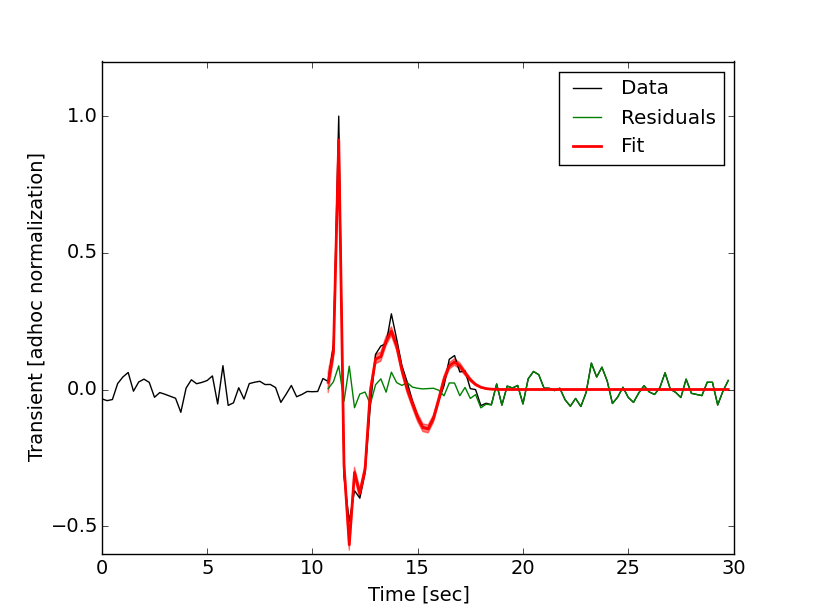}
\caption{\small Gaussian shapelet model of the MICROSCOPE transient whose 1D exponential shapelet model is shown in Fig.\ \ref{fig_crtf}.} \label{fig_gauss}
\end{figure}

The MICROSCOPE transient modelled as exponential shapelets ($n_{\rm max}=15$) in Fig.\ \ref{fig_crtf}, is modelled as Gaussian shapelets ($n_{\rm max}=18$) in Fig.\ \ref{fig_gauss}.
Achieving a similar precision of fit requires more coefficients, and creates more small-scale artefacts: the data are over-fit near the centre of the model, and underfit at the extremes.
We also find that its coefficients are highly covariant, with the good fit produced by large positive and negative basis functions almost precisely cancelling each other. This is much less apparent for the 1D exponential shapelet coefficients.
Indeed, we find that estimating the model again on points different from those observed data fails to reproduce the overall shape of the transient, whereas the 1D exponential shapelet model is itself predictive.

In this example, it appears that 1D exponential shapelets outperform Gaussian shapelets. 
This is because their perturbations around an exponential decay are better matched to the underlying signal, and because their wider extent spreads information density more evenly.
In general, the type of shapelet to choose should depend on the decay rate of the target function.

%%%%%%%%%%%%%%%%%%%%%
%%%%% 2D %%%%%%%%%%%%%%
%%%%%%%%%%%%%%%%%%%%%

\section{2D exponential shapelets} \label{sect_2D_exponential_shapelets}

\begin{figure*}
\centering
\includegraphics[width=\textwidth,angle=0]{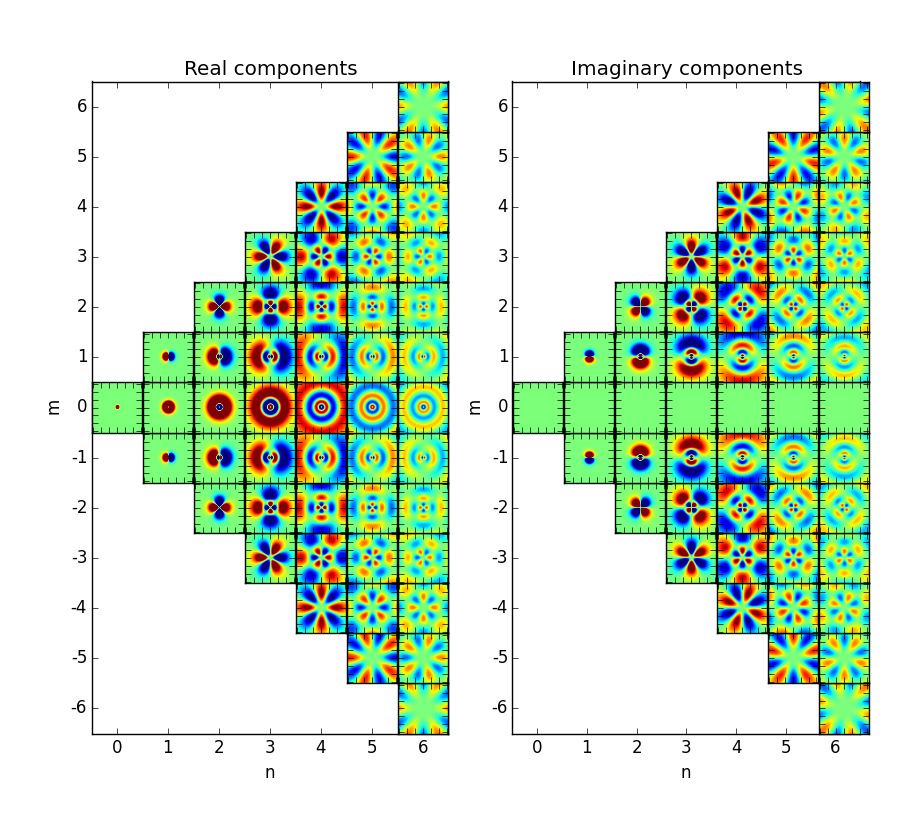}
\caption{\small The first few 2D exponential shapelet basis functions: real part (left) and imaginary part (right). Red is positive, blue is negative. The normalisation of both the colour scale and the spatial scale is arbitrary, but is the same in every box.} \label{fig_xlets}
\end{figure*}

\subsection{Definition}

The quantum mechanics of a hydrogen atom restricted to two dimensions is well studied \citep[see e.g.][]{zz67,yang91,cclk07}.
The bound states of the electron (i.e.\ eigenfunctions of the Hamiltonian) provide a natural set of basis functions to represent a bounded distribution.
After renormalizing these, we define the 2D exponential shapelet basis functions
\begin{multline} \label{eq_psi}
\Psi_{n,m}(r,\phi;\beta) =  (-1)^n \sqrt{\frac{2}{\beta\pi(2n+1)^{3}}\, \frac{(n-|m|)!}{(n+|m|)!}} \\
~~~~~~~~~~~~~~~~~\times \left( \frac{2r}{\beta(2n+1)} \right)^{|m|} L_{n-|m|}^{2|m|}\left( \frac{2r}{\beta(2n+1)} \right) \\
\times
\exp \left(-\frac{r}{\beta(2n+1)} \right) 
\exp(-im\phi),
\end{multline}
for $n\geqslant 0$, where $L_i^j\left(x\right)$ are generalized Laguerre polynomials.
The $(-1)^n$ term is used to ensure that the integral of each basis function is positive, but this is otherwise the form used in quantum theory.
In terms of quantum mechanics, $n$ is the principal quantum number (the eigenfunction energy level) and $m$ the magnetic quantum number, which takes integer values between $-n$ and $n$. The characteristic scale $\beta$ is linked to the Bohr radius.

These functions form an orthonormal set of basis functions in the $L^2([0,\infty[\times [0,2\pi[,\langle\cdot,\cdot\rangle)$ space equipped with inner product $\langle \Psi_{n,m}(r,\phi), \Psi_{n',m'}(r,\phi)\rangle = \int_0^{2\pi} \int_0^\infty \Psi_{n,m}(r,\phi)\, \Psi^*_{n',m'}(r,\phi) \,r\, {\rm d}r\, {\rm d}\phi = \delta_{nn'}\delta_{mm'}$.

Any localized function $f(r,\phi)$ can be uniquely decomposed into a weighted sum of these basis functions
\begin{equation} \label{eq_decomp2}
f(r,\phi)=\sum_{n=0}^\infty \sum_{m=-n}^n f_{n,m}\, \Psi_{n,m}(r,\phi;\beta),
\end{equation}
where the 2D exponential shapelet coefficients are given by
\begin{equation} \label{eq_coeffs}
f_{n,m}=\int_0^{2\pi} \int_0^\infty f(r,\phi)\, \Psi_{n,m}(r,\phi;\beta)\, r\, {\rm d}r\, {\rm d}\phi.
\end{equation}
Using Bessel's inequality like in the 1D case, we find that series (\ref{eq_decomp2}) converges, and the coefficients $f_{n,m}$ must vanish when $n$ and $m$ increase.
For a real function $f(r,\phi)\in\mathbb{R}$, $f_{n,m}=f_{n-m}^*$.
In this case, truncating the series at $n\leqslant n_{\rm max}$ results in $n_{\rm coeffs}=(n_{\rm max}+1)^2$ independent coefficients.

\begin{figure}
\centering
\includegraphics[width=0.5\textwidth,angle=0]{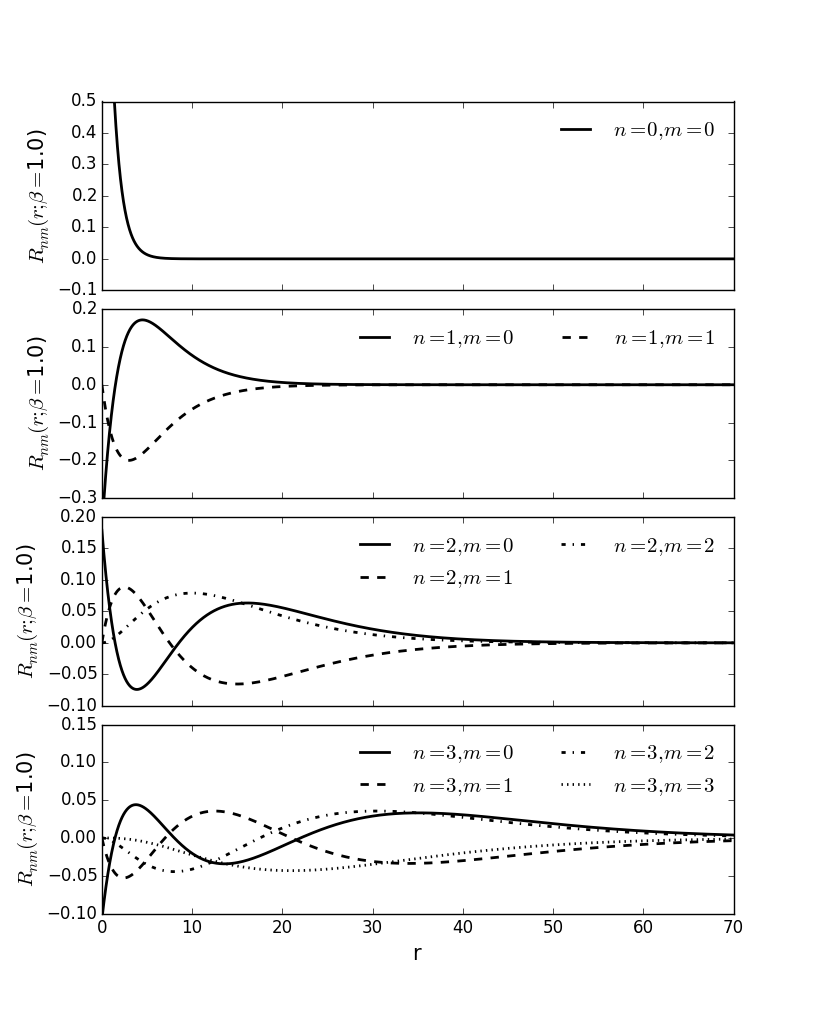} 
\caption{\small Radial component of the first few 2D exponential shapelet basis functions, $R_{n,m}(r)$. Functions with the same $n$ are of the same colour (blue for $n=0$, green for $n=1$, red for $n=2$, black for $n=3$). Functions with $m\neq 0$ are shown as dashed lines.} \label{fig_rnm}
\end{figure}

It may also be possible to define elliptical 2D exponential shapelets by applying a shear transformation to the coordinate system, as \citet{nakajima07} suggested for Gaussian shapelets. This preserves their orthonormality, and can increase their rate of data compression, at the cost of two additional non-linear parameters to specify the shear.

\subsection{Properties}

\subsubsection{Maximum and minimum effective scales}

The first few 2D exponential shapelet basis functions are illustrated in Fig.\ \ref{fig_xlets}, and their radial component is shown in Fig.\ \ref{fig_rnm}, defined via $\Psi_{n,m}(r,\phi;\beta) \equiv R_{n,m}(r;\beta) \exp(-im\phi)$.
Their resemblance with Gaussian polar shapelets is striking \citep[see Fig.\ 2 of][]{polar_shapelets}. 
However, due to their exponential kernel, exponential shapelets are both more peaky and further spread out than Gaussian shapelets. 
The size difference between the lowest order basis function $\Psi_{00}$ and even a low order one like $\Psi_{40}$ is striking. 
It is this behaviour that will help to describe spatially extended features. 

Generalising the 1D case (Section~\ref{sect_1D_theta_minmax}), we define the largest effective scale in a 2D exponential shapelet model as
\begin{eqnarray} 
\theta_{\rm max} &\!\!\approx&\!\! \sqrt{\frac{\iint_{\mathbb{R}^2} r^2\, \Psi_{n_{\rm max},0}(r,\phi;\beta)\,r{\rm d}r{\rm d}\phi}{\iint_{\mathbb{R}^2}  \Psi_{n_{\rm max},0}(r,\phi;\beta)\, r{\rm d}r{\rm d}\phi}} \nonumber \\
&\!\!=&\!\! \beta (2n_{\rm max}+1)\sqrt{2(2n_{\rm max}^2+2n_{\rm max}+3)}\,.
\end{eqnarray}
Again empirically, the smallest resolved scales 
are roughly constant,
\begin{equation}
\theta_{\rm min} \approx \beta.
\end{equation}
The range of scales included in a 2D exponential shapelet model as $n_{\rm max}\rightarrow\infty$ is
$(\theta_{\rm max}/\theta_{\rm min})^2\propto n_{\rm max}^4\propto n_{\rm coeffs}^2$.
The resolution of an exponential shapelet model is greatest near the origin, and information density is concentrated there.
This behaviour is different from the Gaussian shapelets, where resolution is more spatially uniform, and information density is constant, with 
$(\theta_{\rm max}/\theta_{\rm min})^2=n_{\rm max}^2+1\propto n_{\rm coeffs}$.

\subsubsection{Fourier transformation and convolution}

Using the same convention as in the 1D case, it can be shown that the Fourier transform of 2D exponential shapelets is given by
\begin{equation} \label{eq_h2_fourier}
\tilde{\Psi}_{n,m}(\vec{k};\beta) = 2\pi i^m {\rm e}^{-im\phi_k} \left[ F_m(k)\right]_{n,\beta},
\end{equation}
where $k=|\vec{k}|$, $\phi_k$ is the angle between the $\vec{k}$ direction and the $\phi=0$ direction in polar space, and $\left[ F_m(k)\right]_{n,\beta}$ is the Hankel transform of the $R_{nm}(r, \beta)$ radial function.
Consequently, the convolution tensor involved in the convolution of two objects modelled in 2D exponential shapelets is the integral of products of Hankel transforms of $R_{n,m}$ functions. 
We were not able to find an analytic form for this Hankel transform, so it must instead be computed numerically. 

\subsubsection{Coordinate transforms}

It can be useful to know the response of the basis functions to 2D linear coordinate transforms: either to know how to mix the coefficients of a shapelet decomposition in order to perform that transform, or to form combinations of coefficients that are invariant under some transforms.
This was used to construct estimators of the shear distortion applied to images of galaxies by the effect of weak gravitational lensing \citep{shapelets_shear2,shapelets_shear3}. 

A convenient shortcut to calculating those transforms for Gaussian shapelets was provided by the ladder operators associated with the quantum mechanical harmonic oscillator \citep{shapelets2}.
Unfortunately, we have not yet found a useful form of the ladder operators for the 2D hydrogen atom.
If it becomes necessary to perform linear transformations on 2D exponential shapelets, it will be necessary to apply the transforms manually to the basis functions (a long and arduous task, but one that is guaranteed to yield a closed form, because the basis is complete).

\subsection{Object shape measurement}

Once a feature has been decomposed into 2D exponential shapelets, its coefficients can be used to construct characteristic measurements of its shape.
In this section, we derive expressions for an object's (azimuthally-averaged) radial profile, flux, centroid, unweighted quadrupoles, size and ellipticity, in terms of its shapelet coefficients. 
We have not attempted it here, but it should also be possible to develop an exponential-shapelet classification of galaxy morphologies via e.g.\ their concentration, asymmetry and symmetry \citep{cas_morphology}.

\subsubsection{Radial profile}

Azimuthally averaging an object's signal yields its mean radial profile
\begin{equation} \label{eq_profile}
\bar{f}(r)\equiv\frac{1}{2\pi}\int_0^{2\pi}f(r,\phi)\, {\rm d}\phi.
\end{equation}
Noting that all $m\neq 0$ basis functions average to zero, the radial profile reduces to
\begin{equation} \label{eq_profile}
\bar{f}(r)=\sum_{n=0}^\infty f_{n,0} \Psi_{n,0}(r;\beta),
\end{equation}
where the (rotationally invariant) $m=0$ basis functions are
\begin{multline} 
\Psi_{n,0}(r;\beta)=  (-1)^n \frac{2}{\beta \sqrt{2\pi}} (2n+1)^{-3/2} \\
~~~\times \exp\left(-\frac{r}{\beta(2n+1)}\right) L_n\left( \frac{r}{\beta(n+\frac{1}{2})} \right).
\end{multline}
Eq.\ (\ref{eq_profile}) is identical to the equivalent derivation for Gaussian shapelets \citep[see Eq.\ (16) of][]{polar_shapelets}, up to the fact that $m=0$ basis functions with odd $n$ do not exist in the Gaussian case.

\subsubsection{Flux}

Integrating the signal inside a circular aperture of radius $R$ yields its `flux'
\begin{equation}
F_R \equiv \int_0^{2\pi} \int_0^R f(r,\phi)\, r\, {\rm d}r\, {\rm d}\phi.
\end{equation}
To evaluate this integral, it is useful to note that, once again,  all $m\neq 0$ basis functions cancel out to zero during integration over $\phi$, and also a closed form for the generalized Laguerre polynomials,
\begin{equation}
L_n^\alpha(x) = \sum_{k=0}^n (-1)^k \binom{n+\alpha}{n-k} \frac{x^k}{k!}.
\end{equation}
Using this, it can be shown that
\begin{multline}
F_R = 2 \sqrt{2\pi} \beta \sum_{n=0}^\infty f_{n,0}\,(2n+1)^{1/2}  ~~\times \nonumber \\
 \sum_{k=0}^n \frac{2^k (-1)^{n+k}}{k!} \binom{n}{k}  ~ \gamma\left(k+2, \frac{R}{\beta(2n+1)}\right),
\end{multline}
where $\gamma(y,x)$ is the lower incomplete gamma function.
Extrapolating $F_R$ to large radius, and taking the limit $R\rightarrow\infty$, we obtain
\begin{equation} \label{eq_flux}
F \equiv \iint_{\mathbb{R}^2} f(r,\phi)\, r{\rm d}r{\rm d}\phi ~=~ 2\sqrt{2\pi} \beta \sum_{n=0}^\infty (2n+1)^{3/2}f_{n,0}\,.
\end{equation}

\subsubsection{Centroid}

Similarly, it can be shown that the unweighted centroid ($x_c,y_c$), defined by
\begin{equation} \label{eq_centroid}
x_c+i y_c \equiv \frac{\iint_{\mathbb{R}^2}(x+i y) f(x,y)\, {\rm d}x{\rm d}y}{\iint_{\mathbb{R}^2} f(x,y)\, {\rm d}x{\rm d}y},
\end{equation}
is, in terms of 2D exponential shapelet coefficients,
\begin{equation}
x_c+i y_c = -\frac{4\sqrt{2\pi}\beta^2}{F} \sum_{n=1}^\infty \sqrt{n(n+1)(2n+1)^{5}}\,f_{n,1}\, .
\end{equation}

\subsubsection{Quadrupole moments}

The unweighted quadrupole moments of an object
\begin{equation}
F_{ij} = \iint_{\mathbb{R}^2}x_ix_j f(x,y)\, {\rm d}x{\rm d}y
\end{equation}
can be used to define quantities such as its rms size and ellipticity (see below).
They are, in terms of 2D exponential shapelet coefficients,
\begin{multline} \label{eq_quadrupoles11}
F_{11} = 2\sqrt{2\pi} \beta^3 \sum_{n=0}^\infty(2n+1)^{7/2} \\
\times \left[(2n^2+2n+3)f_{n,0} +  2\sqrt{\frac{(n+2)!}{(n-2)!}}\,f_{n,2} \right]
\end{multline}
\begin{multline} \label{eq_quadrupoles22}
F_{22} = 2\sqrt{2\pi} \beta^3 \sum_{n=0}^\infty(2n+1)^{7/2} \\
\times \left[(2n^2+2n+3)f_{n,0} - 2\sqrt{\frac{(n+2)!}{(n-2)!}}\,f_{n,2} \right]
\end{multline}
\begin{equation} \label{eq_quadrupoles12}
F_{12}\! =\! F_{21}\! =\! -4\sqrt{2\pi} \beta^3 i \! \sum_{n=0}^\infty (2n+1)^{7/2}\! \sqrt{\frac{(n+2)!}{(n-2)!}}\,f_{n,2} \,.
\end{equation}

\subsubsection{Size and ellipticity}

Using Eqs.\ (\ref{eq_quadrupoles11})-(\ref{eq_quadrupoles12}), expressions can be obtained to quantify a feature's size 
\begin{eqnarray} \label{eq_size}
R^2 & \equiv & \frac{F_{11}+F_{22}}{F} \nonumber \\
& = & \frac{4\sqrt{2\pi} \beta^3}{F} \sum_{n=0}^\infty  (2n+1)^{7/2}\, (2n^2+2n+3)\,f_{n,0}
\end{eqnarray}
and ellipticity 
\begin{eqnarray} \label{eq_ellipticity}
\varepsilon & \equiv & \frac{F_{11}-F_{22}+2iF_{12}}{F_{11}+F_{22}} \nonumber \\
& = & \frac{8\sqrt{2\pi}\beta^3}{FR^2} \sum_{n=2}^\infty  (2n+1)^{7/2} \sqrt{\frac{(n+2)!}{(n-2)!}}\,f_{n,2}\,.
\end{eqnarray}
In this complex notation with $\varepsilon\equiv |\varepsilon|\cos{(2\phi)}+i |\varepsilon|\sin{(2\phi)}$, $\varepsilon=0$ denotes rotational invariance, and a positive real (imaginary) component denotes elongation along (at $45^\circ$ to) the $x$-axis.

\begin{figure*}
\centering
\includegraphics[width=0.85\textwidth,angle=0]{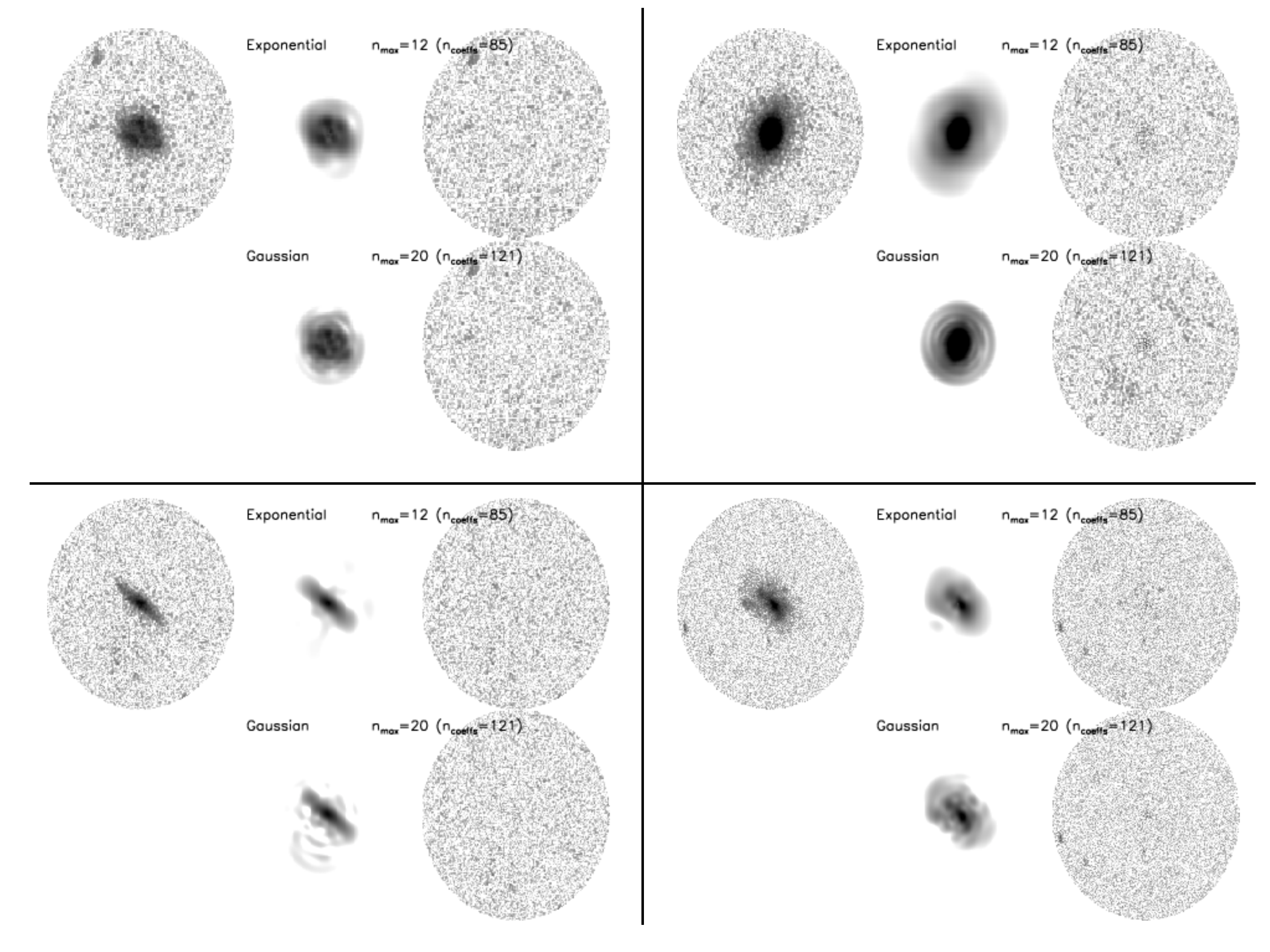}
\caption{\small Test images of four galaxies from the Hubble Space Telescope COSMOS survey \citep{scoville07a, scoville07b}. 
For each galaxy, we show the observed data (upper left); the 2D exponential shapelet model (upper middle) and its residuals (upper right); the Gaussian shapelet model (lower middle) and its residuals (lower right). 
Also shown for each galaxy are the maximum order of decomposition $n_{\rm max}$ and the total number of coefficients $n_{\rm coeffs}$ in each model. 
For a given galaxy, the grey scales are the same in all panels.
} \label{fig_gals}
\end{figure*}

\begin{figure*}
\centering
\includegraphics{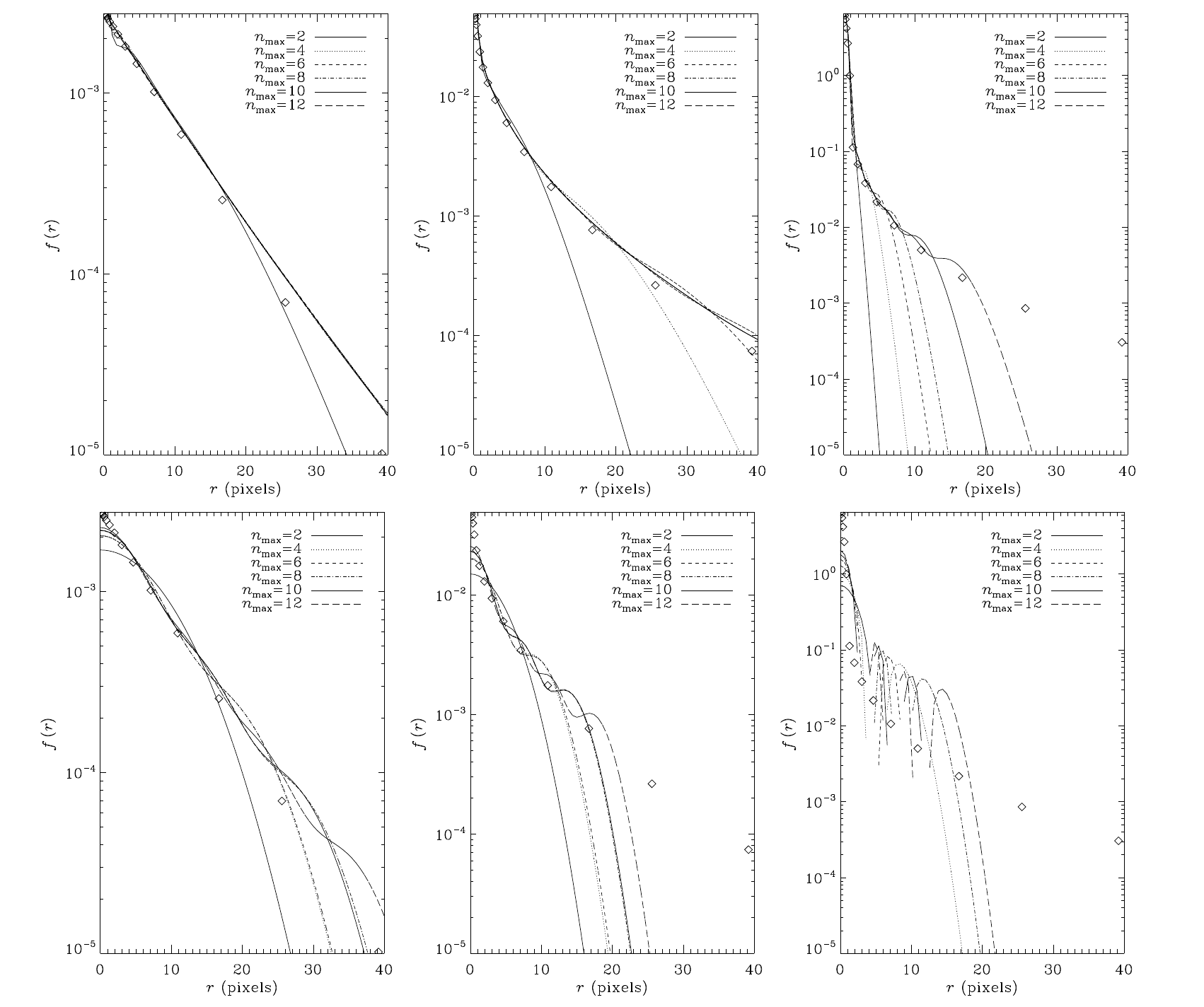}
\caption{\small Exponential (upper panels) and Gaussian (lower panels) shapelets models' speed to converge on the profile of different types of (noiseless) elliptical galaxies: exponential ($n_s=1$, left), intermediate $n_s=2$ (center) and de Vaucouleurs galaxy ($n_s=4$, right). In each panel, squares represent the profile of the galaxy estimated in pixel space.} \label{fig_profs}
\end{figure*}

\begin{figure*}
\centering
\includegraphics{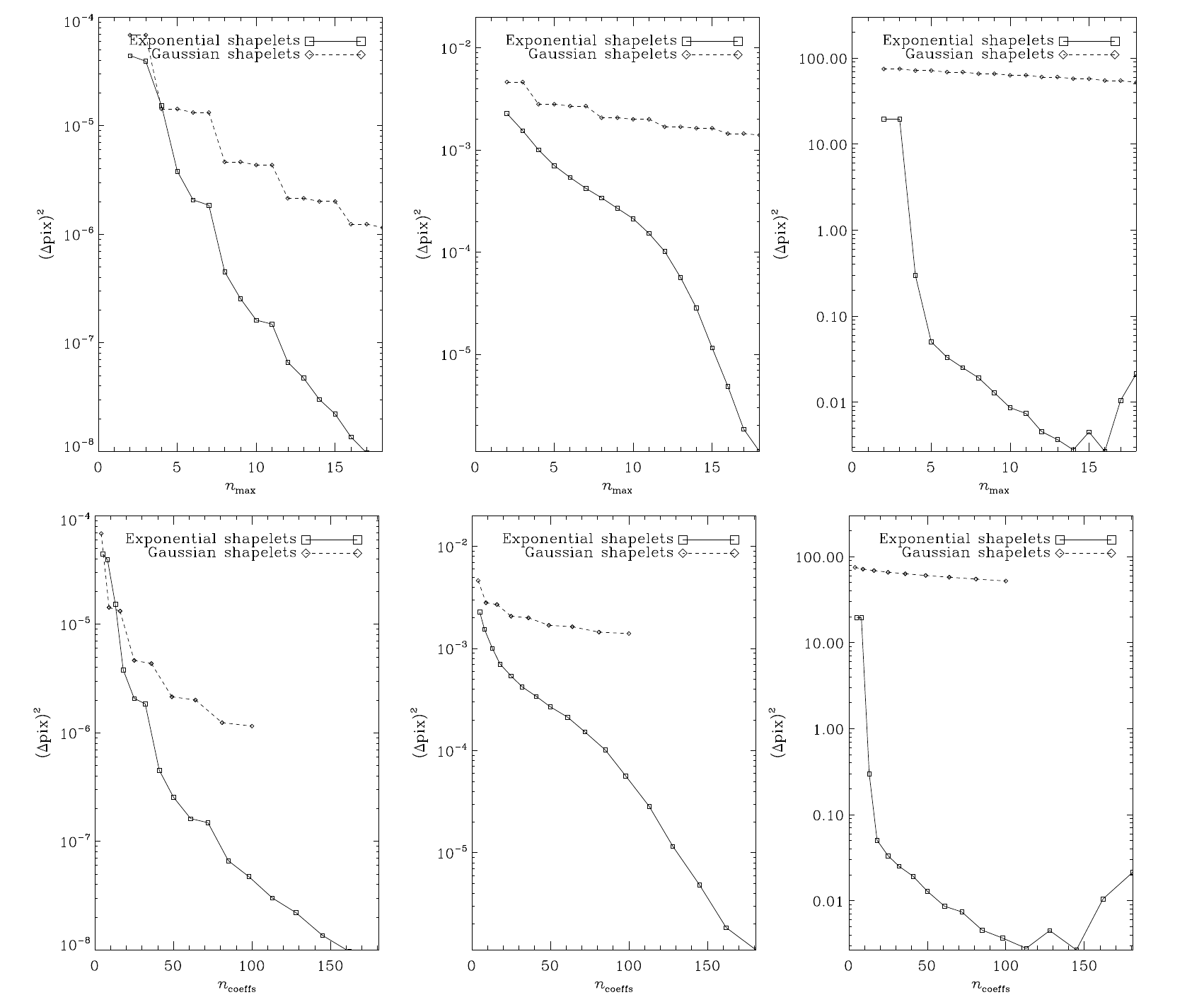}
\caption{\small Convergence speed as a function of shapelets' $n_{\rm max}$ (upper panels) and number of coefficients (lower panels), for exponential (solid lines) and Gaussian (dashed lines) shapelets, for the same Sersic galaxies as in Fig.\ \ref{fig_profs} ($n_s=1,2,4$ from left to right).} \label{fig_convspeed}
\end{figure*}

\subsubsection{Convergence}

Expressions for the shape estimators resemble those obtained for polar Gaussian shapelets in Section~6 of \cite{polar_shapelets}. In particular, a feature's radial profile, flux and size are encoded within the $n=0$ coefficients, while the centroid is described by the $n=1$ coefficients and ellipticity by the $n=2$ coefficients. The expressions also contain the same power of $\beta$ as those for Gaussian shapelets, because that encodes information about the units in which the data is expressed.

However, shape measures for 2D exponential shapelets contain higher powers of $n$ than those for 2D Gaussian shapelets, and will converge more slowly. 
Some of this is due to the normalisation of the basis functions against an inner product. In the Gaussian shapelet case, this happens to yield an integral of basis functions \eqref{eq_flux} that is independent of $n$, while in exponential shapelets it is $\propto n^{3/2}$. That power of $n$ could have been included in the basis functions and removed from the coefficients -- although doing so would merely make them look more stable rather than actually altering convergence.
In the limit that $R\rightarrow\infty$, convergence of the flux estimator requires $|f_{n,0}|$ to decrease faster than $n^{-5/2}$; convergence of centroid requires $|f_{n,1}|$ to decrease faster than $n^{-11/2}$; convergence of size and ellipticity requires $|f_{n,0}|$ and $|f_{n,2}|$ to decrease faster than $n^{-13/2}$.
This may not be a problem, since we expect 2D exponential shapelets to converge faster than shapelets (i.e.\ to have a lower $n_{\rm max}$ for a given galaxy).
However, if this does not hold true, for example due to measurement noise, care should be taken to restrict the decomposition and measurement of photometry inside a finite aperture, or to truncate the 2D exponential shapelet model at finite $n_{\rm max}$.

%%%%%%% Galaxy shape / morphology %%%%%%%%%%%%
\subsection{Example application}

This section demonstrates the use of 2D exponential shapelets to model the shapes of galaxies seen outside our own Milky Way (they may also be convenient to model the shapes of gamma-ray events; S.\ Pires, private communication).
We examine the convergence speed of 2D exponential shapelets, and compare their performance to that of 2D Gaussian shapelets.

\subsubsection{Distant galaxies in astronomical imaging}

Galaxies are collections of (hundreds of billions of) stars, with a combined surface brightness that peaks sharply near the centre, and decreases to large radii in a way that is often exponential.
Four galaxies in the COSMOS survey \citep{scoville07a, scoville07b} observed by the Hubble Space Telescope \citep{cosmos}, are shown in Fig.\ \ref{fig_gals}.
As mentioned in Sect. \ref{ssect_focus} for the 1D case, we follow the algorithm described by \cite{polar_shapelets} to choose the best non-linear parameters of the models --$n_{\rm max}$, $\beta$ and centroid). Note that galaxies in this figure were modelled into shapelets with the ``diamond'' truncation scheme introduced in \cite{shapelets_shear3}, which is intended to reduce small scales oscillations by ignoring the higher $m$ terms of the decomposition. In this case, the number of coefficients is divided by 2, allowing also a better compression of the information.

In all four cases, 2D exponential shapelets require fewer coefficients (lower $n_{\rm max}$; here $n_{\rm coeffs}=85$) to provide a model whose residuals are consistent with the noise than Gaussian shapelets (here $n_{\rm coeffs}=121$). 
Even more interestingly, exponential shapelets tend to provide a (visually) cleaner model of the central region of the galaxies.
The top left galaxy has an irregular morphology, and the 2D exponential shapelet model shows fewer artifacts than the polar shapelet model.
For an elliptical (top right) or edge-on spiral galaxy (bottom left), the ringing common in Gaussian shapelets disappears with 2D exponential shapelets: they do not possess high-frequency ripples at large radii that need large coefficients alternating in sign for them to cancel.
Even for highly eccentric galaxies (bottom right), 2D exponential shapelets provide cleaner models than polar shapelets, especially in the outskirts of the galaxies, as the exponential wings are fundamentally a better match to the underlying signal.

\subsubsection{Convergence speed}

To assess the ability of 2D exponential shapelets to model galaxies, we require noise-free images whose true properties are known.
We simulate elliptical galaxies with a Sersic radial profile
\begin{equation}
\ln\left[\frac{I(r)}{I(r_e)}\right] = -b_n\left[\left(\frac{r}{r_e}\right)^{1/n_s}-1\right],
\end{equation}
where $I(r)$ is the galaxy's s profile, $n_s$ is the Sersic index, $b_n$ is a normalisation constant and $r_e$ is the effective radius (which contains 50\% of the galaxy's flux).
We simulate three types of galaxies: exponential ($n_s=1$), intermediate ($n_s=2$), and de Vaucouleurs ($n_s=4$).

Fig.\ \ref{fig_profs} compares the radial profiles fitted with 2D exponential shapelets (upper row) or Gaussian shapelets (lower row), for different maximum orders of decomposition $n_{\rm max}$, and coarsely optimised $\beta$ (this could be further optimised at low $n_{\rm max}$ in the presence of noise). 
The normalisation of the ordinate depends on $b_n$ and pixellisation, so is arbitrary; only its relative scaling is informative.
For all three galaxy types, 2D exponential shapelets greatly outperform Gaussian shapelets. A decomposition to $n_{\rm max}=2$ or $4$ is sufficient to model an exponential galaxy. Depending on the signal-to-noise of real data, a decomposition to $n_{\rm max}=6$ may be sufficient for an intermediate $n_s=2$ galaxy. 
It is more challenging to model a de Vaucouleurs galaxy, and even exponential shapelets require $n_{\rm max}=12$ to capture its extended tails three orders of magnitude fainter than the peak.
However, they do so without the oscillatory `ringing' introduced by the cancellation of positive and negative basis functions in Gaussian shapelets.

To illustrate the convergence speed, Fig.\ \ref{fig_convspeed} shows the mean square residual of the models of the same galaxies as in Fig.\ \ref{fig_profs},
as a function of the maximum shapelet order $n_{\rm max}$ (upper row), or the total number of coefficients $n_{\rm coeffs}$.
The normalisation of the ordinate is arbitrary, and only its relative scaling is informative.
Once again, it is clear that 2D exponential shapelets (solid lines) outperform Gaussian shapelets (dashed lines). 
For exponential and intermediate galaxies (left and middle columns), the residual of the 2D exponential shapelet model decreases consistently, and 
ends up several orders of magnitude better than that of Gaussian shapelets.
The enhanced performance of exponential shapelets is even more prominent for a de Vaucouleurs galaxy.
The limitation for exponential shapelets in this case is numerical precision of our algorithm to pixellate the basis functions. 
At high-$n$, this breaks down, causing spurious residuals at the very centre of the model, shown as an artificial upturn in the bottom-right panel of Fig.\ \ref{fig_convspeed}.
This could be circumvented by more accurate pixellisation, or by imposing limits on $\theta_{\rm min}$.

%%%%%%% Conclusion %%%%%%%%%%%%
\section{Conclusion} \label{sect_conclusion}

In this paper, we introduced exponential shapelets, a family of orthonormal basis functions that can be efficiently used to model 1D and 2D objects. They borrow many concepts from the original Gaussian shapelets, but are perturbations around a decaying exponential rather than a Gaussian, and more efficiently compress the information in damped 1D oscillations or centrally-concentrated 2D features. In particular, they can simultaneously capture the central peak and the extended wings of galaxies in astronomical imaging, thereby solving the main criticism levelled at Gaussian shapelets in the field for which they were originally intended.

Modelling a feature using exponential shapelets first requires a choice of characteristic scale size $\beta$ and expansion order $n_{\rm max}$; these can be selected using an non-linear optimisation method developed for Gaussian shapelets \citep{polar_shapelets}. The function is then decomposed into a weighted sum of exponential shapelet basis functions using linear regression. A least-squares fit is often sufficient, although in some cases, modeling a transient may be stabilised by regularisation techniques such as sparsity constraints.

Once a feature is described as a weighted sum of exponential shapelets, simple combinations of the weights yield expressions for its total flux, centroid, size and ellipticity. These are similar to those for Gaussian shapelets. However, convolution and deconvolution in exponential shapelets is significantly more difficult than the Gaussian case, with the convolution tensor being time-consuming to calculate via numerical integration.

We described possible applications of exponential shapelets in several fields of experimental and observational science. 
1D exponential shapelets can be used to model (and subtract or understand) spurious transients in time series, such as measurements by the MICROSCOPE or GRACE satellites, or the LIGO search for gravitational waves. 
Thanks to their exponential decay, 2D exponential shapelets overcome the main criticism aimed at Gaussian shapelets (though at the cost of losing some simplicity) and can be used to measure the brightness, shape and size of distant galaxies in astronomical imaging.
The characteristics of exponential shapelets should make them well-suited to data compression and analysis in a wide range of fields; their convenient mathematical properties should see them frequently adopted.

\section*{Acknowledgments}

The concepts in this paper matured over many years, after a preliminary study to generalise Gaussian polar shapelets for Hubble Space Telescope data, for which we acknowledge support from NASA through grant HST-AR-11747 -- and which was carried out at the Jet Propulsion Laboratory, California Institute of Technology, under a contract with NASA. We are indebted to Barnaby Rowe for researching Kummer functions and the literature of the 2D hydrogen atom. We also acknowledge helpful and motivating discussions with Jason Rhodes, Richard Ellis, Alexandre Refregier, Sandrine Pires, James Nightingale, Bernard Foulon, Bruno Christophe, Gilles M\'etris,  Manuel Rodrigues, Alain Robert and Nicolas Touquoy.
This work makes use of technical data from the CNES-ESA-ONERA-CNRS-OCA Microscope mission, and has received financial support from ONERA and CNES.
JB acknowledges the financial support of the UnivEarthS Labex program at Sorbonne Paris Cit\'e (ANR-10-LABX-0023 and ANR-11-IDEX-0005-02) and of CNES through the APR programme (LISA project). RM is supported by a Royal Society University Research Fellowship and through STFC grant ST/N001494/1.

%\section*{References}

\bibliographystyle{mnras}
%\bibliography{bibliography}

\appendix

\onecolumn
\section{Derivation of convolution kernel tensor $C_{nml}$ for 1D exponential shapelets} \label{sect_appA}

\subsection{Proof of Eq. (\ref{eq_cnml})} \label{sect_app1}

Let $f(x)$ and $g(x)$ two 1-dimensional functions, whose convolution product is $h(x) = f(x) \star g(x)$.
In Fourier space, $\tilde{h}(k) = \tilde{f}(k) \tilde{g}(k)$, where
\begin{equation} \label{eq_app_fourier}
\tilde{f}(k) = \frac{1}{\sqrt{2\pi}} \int f(x) e^{ikx} {\rm d}x,
\end{equation}
and similarly for $\tilde{g}(k)$.

Decomposing $f(x)$ and $g(x)$ into exponential shapelets ($f(x) = \sum_{m=1}^\infty f_m \Psi_m(x,\alpha)$), substituting in Eq. (\ref{eq_app_fourier}) and re-arranging terms, we get
\begin{equation} \label{eq_hk}
\tilde{h}(k) = \sum_{m=1}^\infty \sum_{l=1}^\infty f_m g_l \tilde{\Psi}_m(k,\alpha) \tilde{\Psi}_l(k,\beta),
\end{equation}
where we recall that
\begin{equation}
\tilde{\Psi_l}(k,\beta) = (-1)^l \sqrt{\frac{2l\beta}{\pi}} \frac{(l\beta k -i)^{2l}}{\left[ (l\beta k)^2 +1 \right]^{l+1}}.
\end{equation}

Using the Parseval-Plancherel theorem, we get
\begin{equation} \label{eq_hn}
h_n = \int {\rm d}k \tilde{h}(k) \overline{\tilde{\Psi}_n(k,\gamma)},
\end{equation}
where the bar denotes the complex conjugate.

Substituting Eq. (\ref{eq_hk}) in Eq. (\ref{eq_hn}), and re-arranging terms, we find
\begin{equation}
h_n = \sum_{m=1}^\infty \sum_{l=1}^\infty f_m g_l \int_{-\infty}^\infty {\rm d}k (-1)^{n+m+l} \sqrt{\alpha\beta\gamma} ~ \sqrt{mnl} ~
%\times 
\frac{(m\alpha k - i)^{2m}}{[(m\alpha k)^2 + 1]^{m+1}} \frac{(l\beta k - i)^{2l}}{[l\beta k)^2 + 1]^{l+1}} \frac{(n\gamma k - i)^{2n}}{[(n\gamma k)^2 + 1]^{n+1}},
\end{equation}
which defines $C_{nml}$ such that
\begin{equation} \label{eq_app_hn}
h_n = \sum_{m,l=1}^\infty f_m g_l C_{nml}.
\end{equation}
The final expression for $C_{nml}$ (Eq.\ \ref{eq_cnml}) is then found using the binomial decompositions of $(m\alpha k - i)^{2m}$, $(l\beta k - i)^{2l}$ and $(n\gamma k - i)^{2n}$. 

It can noted that $C_{nml}$ is a complex number, unless $u+v+w$ is even, such that $i^{u+v+w}=-1$. If $u+v+w$ is odd, then the integral is 0 (see \ref{sect_app2}). Hence, we do not need to impose any restriction on $u+v+w$ for $C_{nml}$ to be real.
We can also note that since $u \leqslant 2m$, $v \leqslant 2l$ and $w \leqslant 2n$, the integral in Eq. (\ref{eq_cnml}) converges.

We will then aim to look for an analytic expression for the integral in Eq. (\ref{eq_cnml}). We introduce
\begin{equation} \label{eq_app_integral}
\mathcal{I}_{u,v,w}^{m,l,n} \equiv \int_{-\infty}^\infty {\rm d}k (-1)^w \frac{k^{u+v+w}}{\left[(m\alpha k)^2 + 1\right]^{m+1} \left[(l\beta k)^2 + 1\right]^{l+1} \left[(n\gamma k)^2 + 1\right]^{n+1}}
\end{equation}

\subsection{Properties of $\mathcal{I}_{u,v,w}^{m,l,n}$'s integrand} \label{sect_app2}

In this section, we derive some properties of the integrand that appears in the definition of $\mathcal{I}_{u,v,w}^{m, l, n}$. Let us name it $f(x)$ (rigorously, we should write $f_{u,v,w}^{m,l,n}(x)$, but we drop the $u,v,w,n,m,l$ indices to simplify the notation), such that
\begin{equation}
f(x) = (-1)^w \frac{x^{u+v+w}}{\left[(m\alpha x)^2 + 1\right]^{m+1} \left[(l\beta x)^2 + 1\right]^{l+1} \left[(n\gamma x)^2 + 1\right]^{n+1}},
\end{equation}
where $0 \leqslant u \leqslant 2m$, $0 \leqslant v \leqslant 2l$ and $0 \leqslant w \leqslant 2n$ (see Eq.\ \ref{eq_cnml}).

\subsubsection{Alternative definition}
When developping binomial expressions in $f$, we get another form for the function, that appears as the inverse of a series of $x$
\begin{equation}
1/f(x) = (-1)^w \sum_{\nu_m=0}^{m+1} \sum_{\nu_l=0}^{l+1} \sum_{\nu_n=0}^{n+1} \binom{m+1}{\nu_m} \binom{l+1}{\nu_l} \binom{n+1}{\nu_n} (m\alpha)^{2\nu_m} (l\beta)^{2\nu_l} (n\gamma)^{2\nu_n} x^{2(\nu_m + \nu_l + \nu_n) -u-v-w}
\end{equation}

\subsubsection{Parity}
It is obvious that $f(-x) = (-1)^{u+v+w} f(x)$, so that $f$ is even (resp. odd) when $u+v+w$ is even (resp. odd).
Hence, $\mathcal{I}_{u,v,w}^{m,l,n}$ vanishes when $u+v+w$ is odd, in which case $C_{nml}=0$. As mentioned in the main text, $C_{nml}$ is real if $u+v+w$ is even (then, $C_{nml}$ is real for all combinations of $u$, $v$, $w$, $n$, $m$, $l$).
In the remainder of this appendix, we restrict ourselves to $x \geqslant 0$.

\subsubsection{Value in 0}
Evidently, $f(0) = 0$ if $u+v+w \neq 0$. If $u+v+w=0$, then $f(0)=1$.

\subsubsection{Limit at $x \rightarrow \infty$}
Since $0 \leqslant u \leqslant 2m$, $0 \leqslant v \leqslant 2l$ and $0 \leqslant w \leqslant 2n$, it is easy to see that $\lim_{x\rightarrow \infty}=0$.

\subsubsection{Dependence on $n$, $m$, $l$}
For a given $x$, $f$ quickly decreases as $m^{2(m+1)}$ (and similarly for $n$ and $l$), showing that only the contributions of low $n$, $m$ and $l$ are significant in $C_{nml}$. Then, the series $h_n$ (Eq.\ \ref{eq_app_hn}) converges quickly.

\begin{figure*}
\centering
\includegraphics[width=0.45\textwidth,angle=0]{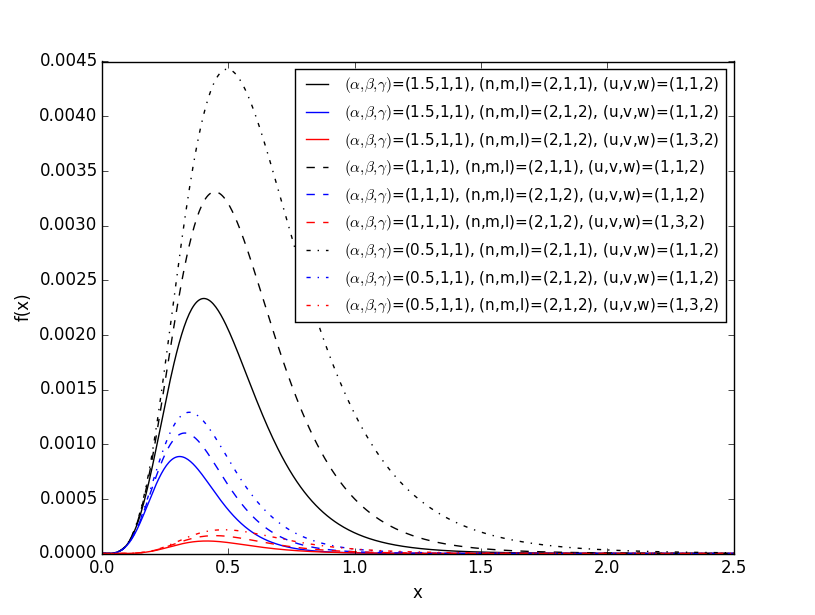}
 \includegraphics[width=0.45\textwidth,angle=0]{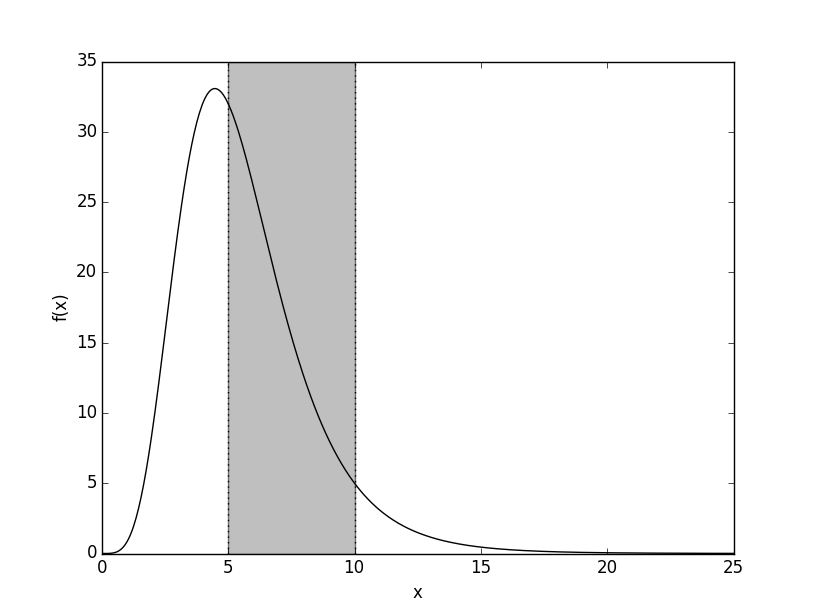}
\caption{\small $f(x)$ function. Left: $f(x)$ for different values of $\alpha$, $\beta$, $\gamma$, $n$, $m$, $l$, $u$, $v$, $w$. Right: $f(x)$ for $\gamma = \alpha = \beta) = 1$, $n$ = 2, $m$ = 1 $l$ = 1, $w$ = 2, $u = 1$, $v = 1$; the grey region shows the domain $[\min \left((m\alpha)^{-1}, (l\beta)^{-1}, (n\gamma)^{-1}\right), \max \left((m\alpha)^{-1}, (l\beta)^{-1}, (n\gamma)^{-1}\right)]$.} \label{fig_app_f}
\end{figure*}

The left panel of Fig.\ \ref{fig_app_f} shows $f(x)$ for some realistic combinations $(\alpha, \beta, \gamma, n, m l, u, v, w)$. It can be seen that $f(x)$ is well-behaved, and tends quickly towards 0 for large $x$. Its extend depends on $\alpha$, $\beta$ and $\gamma$. As mentioned above, for a given $x$, it quickly decreases when $n$, $m$, $l$ increase.
The right panel of Fig.\ \ref{fig_app_f} shows $f(x)$ when  $\gamma = \alpha = \beta) = 0.1$, $n$ = 2, $m$ = 1 $l$ = 1, $w$ = 2, $u = 1$ and $v = 1$, together with the values $x=\min \left((m\alpha)^{-1}, (l\beta)^{-1}, (n\gamma)^{-1}\right)$ and $x=\max \left((m\alpha)^{-1}, (l\beta)^{-1}, (n\gamma)^{-1}\right)$ (borders of the grey area), which are key values in computing $\mathcal{I}_{u,v,w}^{m,l,n}$ (see below).

\subsection{Computation of the integral in $C_{nml}$} \label{sect_app3}

We now turn to look for an analytic expression for $I_{u,v,w}^{m,l,n}$. 
We first note that, due to the parity property of $f_{u,v,w}^{m,l,n}(x)$,
\begin{equation}
\mathcal{I}_{u,v,w}^{m,l,n} \equiv 2 \mathcal{I}_{u,v,w,>}^{m,l,n}= 2 \int_0^\infty f_{u,v,w}^{m,l,n}(k) {\rm d}k
\end{equation}
if $u+v+w$ is even, or $I_{u,v,w}^{m,l,n}=0$ otherwise.

We decompose $I_{u,v,w,>}^{m,l,n}$ as
\begin{equation} \label{eq_int_pos}
\mathcal{I}_{u,v,w,>}^{m,l,n} = \int_0^\mu f_{u,v,w}^{m,l,n}(k) {\rm d}k + \int_\mu^M f_{u,v,w}^{m,l,n}(k) {\rm d}k + \int_M^\infty f_{u,v,w}^{m,l,n}(k) {\rm d}k,
\end{equation}
where, as we show below, we impose $\mu < \min\left((m\alpha)^{-1}, (l\beta)^{-1}, (n\gamma)^{-1}\right)$ and $M > \max\left((m\alpha)^{-1}, (l\beta)^{-1}, (n\gamma)^{-1}\right)$.

\subsubsection{Computation of first integral in r.h.s.\ of Eq.\ (\ref{eq_int_pos})}

We first introduce
\begin{equation} \label{eq_app_integral1}
I_{u,v,w}^{m,l,n}(\mu) \equiv \int_0^\mu f_{u,v,w}^{m,l,n}(k) {\rm d}k,
\end{equation}
where, for clarity, we wrote $f$ with all its indices.

Changing variable such that $k = y\mu$, we get
\begin{equation}
I_{u,v,w, 1}^{m,l,n}(\mu) = (-1)^w \mu^{u+v+w+1} \int_0^1 {\rm d}y \left[(m\alpha \mu)^2y^2+1\right]^{-m-1} \left[(l\beta \mu)^2y^2+1\right]^{-l-1} \left[(n\gamma \mu)^2y^2+1\right]^{-n-1} y^{u+v+w}.
\end{equation}
Another change of variable, $t = y^2$, gives
\begin{equation} \label{eq_app_Iint}
I_{u,v,w}^{m,l,n}(\mu) = \frac{(-1)^w}{2} \mu^{u+v+w+1} \int_0^1 t^{\frac{u+v+w-1}{2}} \left[1+ (m\alpha \mu)^2 t\right]^{-m-1} \left[1+(l\beta \mu)^2 t\right]^{-l-1} \left[1+(n\gamma \mu)^2 t\right]^{-n-1} {\rm d}t,
\end{equation}
where we can recognize, if $\mu < \min\left((m\alpha)^{-1}, (l\beta)^{-1}, (n\gamma)^{-1}\right)$,  Lauricella's function of the fourth kind (e.g. \citealt{bezrodnykh16, hasanov07})
\begin{equation}
F_D^{(r)}(a, b1, \dots, b_r, c ; x_1, \dots, x_r) = \frac{\Gamma(c)}{\Gamma(a) \Gamma(c-a)} \int_0^1 t^{a-1} (1-t)^{c-a-1} \prod_{i=1}^r (1-x_it)^{-b_i} {\rm d}t
\end{equation}
with $r=3$, $a=\frac{u+v+w+1}{2}$, $c = \frac{u+v+w+3}{2}$, $b_1 = m+1$, $b_2 = l+1$, $c_2 = n + 1$, $x_1 = -(m\alpha \mu)^2$, $x_2 = -(l\beta \mu)^2$ and $x_3 = -(n\gamma \mu)^2$, such that
\begin{multline} \label{eq_app_il}
I_{u,v,w}^{m,l,n}(\mu) = \frac{(-1)^w}{2} \mu^{u+v+w+1} \frac{\Gamma\left((u+v+w+1)/2\right)}{\Gamma\left( (u+v+w+3)/2\right)} \\
\times F_D^{(3)}\left( \frac{u+v+w+1}{2}, m+1, l+1, n+1, \frac{u+v+w+3}{2}; -(m\alpha L)^2, -(l\beta \mu)^2, -(n\gamma L)^2 \right).
\end{multline}

$F_D^{(3)}$ is defined (as long as $\max(|x_i|)<1$) as a converging series (Eq. (1.4) in \citealt{hasanov07}), such that
\begin{equation}
F_D^{(r)}(a, b1, \dots, b_r, c ; x_1, \dots, x_r) = \sum_{\nu1, \dots, \nu_r=0}^\infty \frac{(a)_{\nu_1 + \dots + \nu_r} (b_1)_{\nu_1} \dots (b_r)_{\nu_r}}{(c)_{\nu_1+\dots+\nu_r}} \frac{x_1^{\nu_1}}{\nu_1!}\dots \frac{x_r^{\nu_r}}{\nu_r!},
\end{equation}
where $(a)_\nu$ is the Pochhammer symbol, such that (after simplifying Pochhammer symbols):
\begin{multline} \label{eq_app_I}
I_{u,v,w}^{m,l,n}(\mu) = (-1)^w \sum_{\nu_1,\nu_2,\nu_3=0}^\infty \frac{(-1)^{\nu_1+\nu_2+\nu_3}}{u+v+w+2(\nu_1+\nu_2+\nu_3) + 1} \frac{(m+1)_{\nu_1} (l+1)_{\nu_2} (n+1)_{\nu_3}}{\nu_1!\nu_2!\nu_3!} \\
\times (m\alpha)^{2\nu_1} (l\beta)^{2\nu_2} (n\gamma)^{2\nu_3} \mu^{u+v+w+2(\nu_1+\nu_2+\nu_3)}
\end{multline}

\subsubsection{Computation of second integral in r.h.s.\ of Eq.\ (\ref{eq_int_pos})}

We then introduce
\begin{equation} \label{eq_app_integral2}
J_{u,v,w}^{m,l,n}(\mu,M) \equiv \int_\mu^M f_{u,v,w}^{m,l,n}(k) {\rm d}k,
\end{equation}
that we will compute in a similar fashion as the previous integral.

With successive changes of variables $y=k^2$, $z=y-\mu^2$, and $t=z/(M^2-\mu^2)$ (under our assumptions, $M \neq \mu$; if $M=\mu$, $J_{u,v,w}^{m,l,n}(\mu,M)=0$), we obtain
\begin{multline} \label{eq_app_Jint}
J_{u,v,w}^{m,l,n}(\mu,M) = \frac{(-1)^w (M^2-\mu^2) \mu^{u+v+w-1}}{2 \left[ 1 + (\mu m\alpha)^2\right]^{m+1} \left[ 1 + (\mu l\beta)^2\right]^{l+1} \left[ 1 + (\mu n\gamma)^2\right]^{n+1}} \\
\times \int_0^1 \left[ 1 - \frac{\mu^2-M^2}{\mu^2}t\right]^{\frac{u+v+w-1}{2}} \left[1 - \frac{(\mu^2-M^2)(m\alpha)^2}{1+(\mu m\alpha)}t \right]^{-m-1} 
\left[1 - \frac{(\mu^2-M^2)(l\beta)^2}{1+(\mu l\beta)}t \right]^{-l-1} \left[1 - \frac{(\mu^2-M^2)(n\gamma)^2}{1+(\mu n\gamma)}t \right]^{-n-1} {\rm d}t
\end{multline}

If $M \geqslant \sqrt{2}\mu$, then $J_{u,v,w}^{m,l,n}(\mu,M)$ must be calculated numerically. This is the case in the example of Fig.\ \ref{fig_app_f}. 
However, if
%If 
$M<\sqrt{2}\mu$ (i.e.\, $\max\left\{(m\alpha)^{-1}, (l\beta)^{-1}, (n\gamma)^{-1}\right\}$ $<$ $\sqrt{2} \min\left\{(m\alpha)^{-1}, (l\beta)^{-1}, (n\gamma)^{-1}\right\}$, then we recognize a Lauricella $F_D^{(4)}$ function, such that
\begin{multline}
J_{u,v,w}^{m,l,n}(\mu,M) = \frac{(-1)^w (M^2-\mu^2) \mu^{u+v+w-1}}{2 \Gamma(2) \left[ 1 + (\mu m\alpha)^2\right]^{m+1} \left[ 1 + (\mu l\beta)^2\right]^{l+1} \left[ 1 + (\mu n\gamma)^2\right]^{n+1}} \\
\times F_D^{(4)} \left(1, \frac{1-u-v-w}{2}, m+1, l+1, n+1, 2; \frac{\mu^2-M^2}{\mu_2}, \frac{(\mu^2-M^2)(m\alpha)^2}{1+(\mu m\alpha)^2}, \frac{(\mu^2-M^2)(l\beta)^2}{1+(\mu l\beta)^2} \frac{(\mu^2-M^2)(n\gamma)^2}{1+(\mu n\gamma)^2}\right).
\end{multline}
If $M<\sqrt{2}\mu$, $J_{u,v,w}^{m,l,n}(\mu,M)$ can finally be expressed as a converging series
\begin{multline} \label{eq_app_J}
J_{u,v,w}^{m,l,n}(\mu,M) = \frac{(-1)^{w+1}}{2} \sum_{\nu_1,\nu_2,\nu_3, \nu_4=0}^\infty \frac{\left( \frac{1-u-v-w}{2}\right)_{\nu_1} (m+1)_{\nu_2} (l+1)_{\nu_3} (n+1)_{\nu_4}}{(\nu_1+\nu_2+\nu_3+\nu_4) \nu_1! \nu_2! \nu_3! \nu_4!} \\
\times \frac{\mu^{-2\nu_1} (\mu^2-M2)^{\nu_1+\nu_2+\nu_3+\nu_4} (m\alpha)^{2\nu_2} (l\beta)^{2\nu_3} (n\gamma)^{2\nu_4}}{\left[ 1+ (\mu m\alpha)^2\right]^{m+1+\nu_2} \left[ 1+ (\mu l\beta)^2\right]^{l+1+\nu_3} \left[ 1+ (\mu n\gamma)^2\right]^{n+1+\nu_4}}.
\end{multline}

\subsubsection{Computation of third integral in r.h.s of Eq. (\ref{eq_int_pos})}

We now introduce
\begin{equation} \label{eq_app_integral3}
K_{u,v,w}^{m,l,n}(M) \equiv \int_M^\infty f_{u,v,w}^{m,l,n}(k) {\rm d}k,
\end{equation}
that we will compute in a similar fashion as $I_{u,v,w}^{m,l,n}$ and $J_{u,v,w}^{m,l,n}$.

We first note that $\int_M^\infty f_{u,v,w}^{m,l,n}(k) {\rm d}k = \lim_{\xi\rightarrow \infty} I_\xi(M)$, where for simplicity, we droped the indices from the $K_\xi$ definition, and $\xi$ is some positive cut-off, and
\begin{equation}
K_\xi \equiv  \int_M^\xi f_{u,v,w}^{m,l,n}(k) {\rm d}k
\end{equation}
$K_\xi$ can be computed using successive changes of variable.
First, we set $y=1/k$, so that (after rearranging some terms)
\begin{equation}
K_\xi = \int_{1/\xi}^{1/M} \frac{(-1)^w y^{-2-(u+v+w)-2(3+n+m+l)}}{(m\alpha)^{2(m+1)}(l\beta)^{2(l+1)}(n\gamma)^{2(n+1)}} \left[ 1 + \frac{y^2}{(m\alpha)^2}\right]^{-m-1} \left[ 1 + \frac{y^2}{(l\beta)^2}\right]^{-l-1} \left[ 1 + \frac{y^2}{(n\gamma)^2}\right]^{-n-1} {\rm d}y.
\end{equation}
Additional changes of variables ($z=y^2$, $t'=z-1/\xi^2$, $t= \frac{M^2\xi^2}{\xi^2-M^2}t$) then yield
\begin{multline}
K_\xi = \frac{(-1^w)}{2(m\alpha)^{2(m+1)}(l\beta)^{2(l+1)}(n\gamma)^{2(n+1)}} \frac{\xi^2-M^2}{\xi^2M^2} \left( \frac{\xi^2 (m\alpha)^2+1}{\xi^2(m\alpha)^2}\right)^{-m-1} \left( \frac{\xi^2 (l\beta)^2+1}{\xi^2(l\beta)^2}\right)^{-l-1} \left( \frac{\xi^2 (n\gamma)^2+1}{\xi^2(n\gamma)^2}\right)^{-n-1} \\
\times \int_0^1 \left( \frac{1}{\xi^2} + \frac{\xi^2-M^2}{\xi^2M^2}t\right)^g \left(1 + \frac{\xi^2-M^2}{M^2[(\xi m\alpha)^2-1]}t\right)^{-m-1} \left(1 + \frac{\xi^2-M^2}{M^2[(\xi l\beta)^2-1]}t\right)^{-l-1} \left(1 + \frac{\xi^2-M^2}{M^2[(\xi n\gamma)^2-1]}t\right)^{-n-1} {\rm d}t
\end{multline}
where $g=3+n+m+l-(3+u+v+w)/2$.
Taking the limit $\xi \rightarrow \infty$, we obtain
\begin{multline} \label{eq_app_Kint}
K_{u,v,w}^{m,l,n}(M) = \frac{(-1)^w M^{-2(3+n+m+l)+1+u+v+w}}{2(m\alpha)^{2(m+1)}(l\beta)^{2(l+1)}(n\gamma)^{2(n+1)}} \\
\times \int_0^1 t^{3+n+m+l-(3+u+v+w)/2} \left[1+\frac{t}{(Mm\alpha)^2}\right]^{-m-1} \left[1+\frac{t}{(Ml\beta)^2}\right]^{-l-1} \left[1+\frac{t}{(Mn\gamma)^2}\right]^{-n-1} {\rm d}t
\end{multline}

Under the assumption that $M > \max\left((m\alpha)^{-1}, (l\beta)^{-1}, (n\gamma)^{-1}\right)$, we recognize a Lauricella $F_D^{(3)}$ function, such that
\begin{multline}
K_{u,v,w}^{m,l,n}(M) = \frac{(-1)^w M^{-2(3+n+m+l)+1+u+v+w}}{2(m\alpha)^{2(m+1)}(l\beta)^{2(l+1)}(n\gamma)^{2(n+1)}} \frac{\Gamma(n+m+l-\frac{u+v+w+5}{2})}{\Gamma(n+m+l-\frac{u+v+w+7}{2})} \\
\times F_D^{(3)} \left(n+m+l-\frac{u+v+w+5}{2}, m+1, l+1, n+1, n+m+l-\frac{u+v+w+7}{2}; -(Mm\alpha)^{-2}, -(Ml\beta)^{-2}, -(Mn\gamma)^{-2}\right).
\end{multline}
and we can finally express $K_{u,v,w}^{m,l,n}(M)$ as a converging series
\begin{multline} \label{eq_app_K}
K_{u,v,w}^{m,l,n}(M) = (-1)^w \sum_{\nu_1,\nu_2,\nu_3=0}^{\infty} \frac{(-1)^{\nu_1+\nu_2+\nu_3}}{5+2(n+m+l)-(u+v+w)} \frac{(m+1)_{\nu_1} (l+1)_{\nu_2} (n+1)_{\nu_3}}{\nu_1!\nu_2!\nu_3!} \\
\times (m\alpha)^{-2(m+1-\nu_1)} (l\beta)^{-2(l+1-\nu_2)} (n\gamma)^{-2(n+1-\nu_3)} M^{-2(3+n+m+l)+u+v+w+1-2(\nu_1+\nu_2+\nu_3)}.
\end{multline}

\bsp
\label{lastpage}

\end{document}